\documentclass[aps,prx,nofootinbib]{revtex4}

\usepackage[english]{babel}
\usepackage{amsfonts,amssymb,amsmath}
\usepackage{physics,mathrsfs,times,multirow}
\usepackage{graphicx} 
\usepackage{amsbsy}
\usepackage{amssymb}
\usepackage{amsthm}
\usepackage{float}

\usepackage[usenames, dvipsnames]{color} 
\usepackage{xcolor}

\usepackage{hyperref}
\usepackage{mathtools}

\usepackage{placeins} 

\definecolor{pdcolor}{rgb}{1,0.5,0}

\definecolor{pdblue}{rgb}{0,0,1}

\definecolor{rkgreen}{rgb}{0,1,0}

\newcommand{\N}{\mathbb{N}}
\newcommand{\Z}{\mathbb{Z}}
\newcommand{\Exp}{\mathbb{E}}

\newcommand{\Orb}{\mathcal{O}}

\newcommand{\im}{\mathrm{i}}

\newcommand{\Pks}{\hat{\tilde{P}}(k,s)}

\begin{document}

\preprint{}

\title{Singularity of L\'evy walks in the lifted Pomeau-Manneville map}

\author{Samuel Brevitt}
\email{s.brevitt@qmul.ac.uk}
\affiliation{Centre for Complex Systems, School of Mathematical
  Sciences, Queen Mary University of London, Mile End Road, London E1
  4NS, UK}

\author{Alexander Schulz}
\email{alexander.schulz2@htw-dresden.de}
\affiliation{Hochschule für Technik und Wirtschaft Dresden,
  Friedrich-List-Platz 1, 01069 Dresden, Germany}

\author{Dominic Pegler}
\email{ecdjpegler@ust.hk}
\affiliation{Department of Economics, HKUST Business School, The Hong
  Kong University of Science and Technology, Clear Water Bay, Kowloon,
  W6 8RH, Hong Kong}

\author{Holger Kantz}
\email{kantz@pks.mpg.de}
\affiliation{Max Planck Institute for the Physics of Complex Systems,
  N\"othnitzer Stra{\ss}e 38, 01187 Dresden, Germany}

\author{Rainer Klages}
\email{r.klages@qmul.ac.uk}
\affiliation{Centre for Complex Systems, School of Mathematical
  Sciences, Queen Mary University of London, Mile End Road, London E1
  4NS, United Kingdom}
\affiliation{London Mathematical Laboratory, 8 Margravine Gardens,
  London W6 8RH, United Kingdom}

\date{\today}

\begin{abstract}

Since groundbreaking works in the 1980s it is well-known that simple
deterministic dynamical systems can display intermittent dynamics and
weak chaos leading to anomalous diffusion. A paradigmatic example is
the Pomeau-Manneville (PM) map which, suitably lifted onto the whole
real line, was shown to generate superdiffusion that can be reproduced
by stochastic L\'evy walks (LWs). Here we report that this matching
only holds for parameter values of the PM map that are of Lebesgue
measure zero in its two-dimensional parameter space. This is due to a
bifurcation scenario that the map exhibits under variation of one
parameter. Constraining this parameter to specific singular values at
which the map generates superdiffusion by varying the second one, as
has been done in previous literature, we find quantitative deviations
between deterministic diffusion and diffusion generated by stochastic
LWs in a particular range of parameter values, which cannot be cured
by simple LW modifcations. We also explore the effect of aging on
superdiffusion in the PM map and show that this yields a profound
change of the diffusive properties under variation of the aging time,
which should be important for experiments. Our findings demonstrate
that even in this simplest well-studied setting, a matching of
deterministic and stochastic diffusive properties is non-trivial.
  
\end{abstract}


\maketitle

\begin{quotation}

  {\bf L\'evy walks are pioneering, very popular models in stochastic
    theory that have been widely used to understand diffusive
    spreading in many different fields of science. They provide a
    basic mechanism to generate anomalous superdiffusion, thus
    defining a complete different class of stochastic processes
    compared to normal diffusive Brownian motion. L\'evy walks were
    first applied to understand superdiffusion in a simple weakly
    chaotic dynamical system that yields a deterministc random walk on
    the line. Here, we show that this matching between deterministic
    and stochastic diffusion breaks down for typical values of the two
    control parameters in the original model. This is due to a
    bifurcation scenario that defies any mapping onto a simple
    stochastic process.  We also report more subtle quantitative
    deviations even in the singular case of parameter values where the
    deterministic system does generate superdiffusion.}
  
\end{quotation}
  

\section{Introduction\label{sec:intro}}

Diffusion is a fundamental feature of many systems in nature,
technology and society \cite{BCKV18}. Many diffusive processes, such
as Brownian motion, are classically characterised by the mean square
displacement (MSD) of a diffusing particle increasing linearly with
time, $\langle x^2 \rangle \sim t$. However it is also more recently
observed in many natural processes that the MSD may instead increase
as $t^\beta$, where $\beta$ can be lesser (called \emph{subdiffusion})
or greater (\emph{superdiffusion}) than unity (\emph{normal
diffusion}). Observations of anomalous diffusion are found in fields
as diverse as physics, microbiology, chemistry and finance
\cite{BoGe90,CKW04,KRS08,metzler_random_2000,metzler_anomalous_2014,
HoFr13,zaburdaev_levy_2015}.

Independently of this, the development of chaos theory led to an
interest in the pseudo-stochastic and diffusive properties of
deterministic chaotic dynamical systems
\cite{EvMo90,HoB99,gaspard_chaos_1998,
  dorfman_introduction_1999,Zas02, klages_microscopic_2007,CFLV08}.
The simplest examples are given by deterministically chaotic one
dimensional maps, introduced and studied in pioneering works at the
beginning of the 1980s \cite{GF2,SFK,GeNi82}, whose trajectories have
a natural relationship to those of random walks. In the decades since,
their diffusive dynamics have been explored by many different methods,
such as stochastic theory
\cite{GF2,SFK,GeNi82,GeTo84,geisel_accelerated_1985,ShlKl85,
  zumofen_scale-invariant_1993,ZuKl93b,klages_simple_1995,dcrc,
  BaKl97,klages_simple_1999,klages_understanding_2002,Bar03,
  KCKSG06,korabel_fractal_2007,AMHR22}, periodic orbit theory
\cite{Art1,ACL93,DeCvi97,DeDa98,CAMTV01}, thermodynamic formalism
\cite{WaHu93,SSR95} and transfer operator techniques
\cite{klages_simple_1995,klages_simple_1999,GrKl02,TG04}.  These maps
were found to exhibit many interesting properties, like fractal
parameter dependencies of transport coefficients
\cite{klages_simple_1995,GrKl02,KCKSG06,korabel_fractal_2007,
  klages_microscopic_2007}, transitions to weak chaos (in terms of
sub-exponential separation of trajectories)
\cite{gaspard_sporadicity_1988,ZaUs01,klages_weak_2013} and
intermittency (periods of chaos punctuated by periods of laminar flow)
\cite{manneville_intermittency_1979,manneville_intermittency_1980,
  pomeau_intermittent_1980} yielding different types of diffusion,
which make them useful models for a variety of systems observed in the
natural sciences.

One important class of low-dimensional, time-discrete maps is the
Pomeau-Manneville (PM) map, first constructed in order to study the
chaos observed in turbulent flows in continuous-time Lorenz systems
\cite{pomeau_intermittent_1980,manneville_intermittency_1980,
  manneville_intermittency_1979}.
The map may be defined {in its reduced form on the unit interval} by
\begin{equation} \label{eqn:reduced}
  x_{n+1} := \tilde{M}(x_n) \quad \text{for} \quad \tilde{M}(x) := x + ax^z \ \text{(mod $1$)}
\end{equation}
depending on parameters $a>0$ and $z\geq 1$ for $0\leq x\leq1$,
where $x_n$ is the position of a point at discrete time $n\in\mathbb{N}$.
This map has an unstable marginal fixed point at $x=0$ which for
certain ranges of the nonlinearity parameter $z$ is known to
produce intermittency and weak chaos
\cite{manneville_intermittency_1979,gaspard_sporadicity_1988}.
{By imposing symmetry and translation relations,
this map can be extended across the real line,
to produce a map exhibiting either subdiffusion or superdiffusion
by careful positioning of the fixed point
\cite{GeTo84,zumofen_scale-invariant_1993,WaHu93,ACL93,korabel_fractal_2007};}
{see Sec.~\ref{sec:setup} for details.}

For the subdiffusive extension of the PM map \eqref{eqn:reduced}, the
time-scaling of the MSD and its dependence on the nonlinearity
parameter $z$ have been determined by different methods
\cite{GeTo84,ACL93,zumofen_scale-invariant_1993,ZuKl93b,WaHu93,
  DeCvi97,BaKl97,DeDa98,Bar03,KCKSG06,korabel_fractal_2007}, as has
the associated generalised diffusion coefficient (GDC)
\cite{DeCvi97,DeDa98}, which is the multiplicative constant in front
of the MSD
\cite{metzler_random_2000,korabel_fractal_2007,klages_weak_2013}\footnotemark
{
\begin{equation}\label{eq:gdc}
K := \lim_{t\to\infty} \frac{\langle x^2 (t) \rangle}{t^\beta}.
\end{equation}}
Its full dependence on the two control parameters $a$ and $z$ was
reported in \cite{KCKSG06,korabel_fractal_2007}, where computer
simulation results were compared with analytic approximations obtained
from continuous-time random walk (CTRW) theory. It was shown that
under variation of $z$ the GDC becomes zero right at the
transition from normal to subdiffusion.

\footnotetext{It should be noted that the definition of the GDC in
  \eqref{eq:gdc} differs from the standard definition of the
  `classical' diffusion coefficient $D := \lim_{t\to\infty}
  \frac{\langle x^2 \rangle}{2t}$ commonly used in physics. Here we
  apply the convention as defined in \eqref{eq:gdc} as introduced in
  relation to anomalous diffusion \cite{metzler_random_2000}.}

For the superdiffusively extended PM map, similar studies {to
  those above were performed} on the special case of parameter values
$a=2^z$, \cite{zumofen_scale-invariant_1993,ZuKl93b,WaHu93,ACL93,
  geisel_accelerated_1985,ShlKl85,TG04}. They revealed that as the
nonlinearity parameter $z$ increases, the MSD transitions through
three different regimes: from normal diffusion ($\beta=1$) to
superdiffusion ($1<\beta<2$) to ballistic motion ($\beta=2$). Most
well-known became the groundbreaking works by Geisel et
al.\ \cite{geisel_accelerated_1985}, and Shlesinger, Zumofen and
Klafter \cite{zumofen_scale-invariant_1993,ZuKl93b,ShlKl85}. They led
to the formulation and first application of the {\em L\'evy walk}
(LW), a fundamental superdiffusive process in stochastic theory,
introduced in parallel purely on the basis of stochastic CTRW theory
\cite{SKW82,KBS87,Shles87}. Due to wide applications across many
different fields of science LWs became quite famous and popular over
the past few decades; see \cite{zaburdaev_levy_2015} for a review. The
GDC associated to LWs was calculated by CTRW theory \cite{ZuKl93b} and
is recovered as a special case from related, more general CTRW walk
models studied very recently
\cite{zaburdaev_superdiffusive_2016,albers_exact_2018,
  albers_nonergodicity_2022,bothe_mean_2019}.  Note, however, that
CTRW theory represents {only} a stochastic modelling of the map's
original deterministic dynamics by using simplifying assumptions.
Hence, in contrast to the calculations in \cite{WaHu93,ACL93,TG04}
based on methods of dynamical systems theory, applying stochastic LW
results to deterministic maps provides no derivation of their
diffusive properties from first principles. {Furthermore,} and
for the work presented here most importantly, so far {MSD and GDC
  have only been calculated for} parameter values under the constraint
{$a=2^z$} for which the PM map exhibits full branches on unit
intervals, which in turn generates particularly simple dynamics
\cite{klages_simple_1995,klages_simple_1999}. Without this constraint,
ie., under independent variation of both $a$ and $z$, the GDC of the
superdiffusive PM map has been investigated numerically only very
recently \cite{pegler_anomalous_2017,schulz_parameter-dependent_2020}
and not been calculated analytically. In particular, {the}
well-known {stochastic } LW results for MSD and GDC have not been
compared with results for the PM map under variation of both
parameters.
References~\cite{pegler_anomalous_2017,schulz_parameter-dependent_2020}
reported preliminary simulation results for the GDC under variation of
the {nonlinearity} $z$, while
\cite{schulz_parameter-dependent_2020} focused on the dependence of
the GDC related to the parameter $a$.

Exact analytical results for transport coefficients of dynamical
systems, like the GDC, under general parameter variation have been
obtained for simple piecewise linear maps (see, eg.,
\cite{klages_microscopic_2007} for a review). But to our knowledge,
there are none for nonlinear dynamical systems like the PM map. A
common approach to understand diffusion in these systems is thus to
compare computer simulation results for the deterministic dynamics
with approximations from stochastic theory, as discussed above. This
approach, however, is generally non-trivial, since {\em per se} there
is no reason why stochastic modelling should yield exact results for a
given deterministic dynamical system. There is a wide range of
examples illustrating this difficulty, eg., diffusion generated by
simple one-dimensional maps (normal \cite{klages_simple_1995}, sub-
\cite{korabel_fractal_2007} and superdiffusion
\cite{pegler_anomalous_2017,schulz_parameter-dependent_2020}),
nonlinear one-dimensional maps exhibiting bifurcation scenarios
\cite{korabel_fractal_2002,korabel_fractality_2004}, two-dimensional
standard maps \cite{ReWi80}, polygonal billiards
\cite{klages_microscopic_2007}, periodic Lorentz gases
\cite{MaZw83,KlDe00,GilSan10}, and soft Lorentz gases
\cite{klages_normal_2019}, to name a few. We furthermore remark that
transport coefficients in low-dimensional periodic deterministic
dynamical systems typically exhibit fractal parameter dependencies,
due to topological instability and periodic orbits
\cite{klages_microscopic_2007}, which defies a naïve understanding in
terms of stochastic theory. Random walk dynamics of these systems is
only recovered in certain special \cite{klages_simple_1999} or
limiting \cite{dcrc} cases of parameter values which allow some
mapping onto a Markov process.

Accordingly, in this article we investigate to what extent the MSD of
the superdiffusively-extended PM map, considered as a function of its
{parameters $z$ and $a$,} can be understood in terms of a
corresponding stochastic LW model. Our two main results are, first,
that the special setting $a=2^z$, which traditionally has been studied
in the literature, is the only one in parameter space that yields
dynamics compatible with LWs; and second, that even in this setting,
in a certain range of parameter values there are subtle deviations
between the GDC calculated from stochastic LW theory and the one
obtained numerically from the deterministic PM map. Our paper is
organised as follows: After introducing the superdiffusive extension
of the PM map at the beginning of Sec.~\ref{sec:setup}, we briefly
outline the methods used for our numerical analysis
(Sec.~\ref{sec:meth}) before we review the corresponding stochastic LW
model formulated in terms of CTRW theory
{(Sec.~\ref{sec:ctrw}). In Sec.~\ref{sec:newsec} we describe and
  discuss the numerical results obtained under variation of a more
  conveniently chosen parameter related to $a$, and demonstrate that
  superdiffusion is only attained for the particular singular
  parameter values $a=2^z$, while the dynamics are either normal or
  localised for others; this is related to a fractal bifurcation
  scenario, which we investigate in this section.}  In
Sec.~\ref{sec:results} {we give numerical results for the GDC of
  the map under variation of its nonlinearity parameter $z$, fixing
  $a=2^z$, as has been done before, and compare these to those
  predicted by LW theory.}  We provide a detailed analysis of the
different diffusive regimes, and the transitions between them, in
terms of the GDC.  {We discover a noteworthy discrepancy between
  the PM map and LWs in the regime of normal diffusion, which we
  investigate and partially explain. Finally,} in Sec.~\ref{sec:aging}
we explore the effect of aging on the GDC and compare our simulations
with recently reported analytic results calculated from an aged LW.
We summarise in Sec.~\ref{sec:concl}, and discuss some potential
physical applications of this work.

\section{Model and methods\label{sec:setup}}

\begin{figure}[t]
\centering
\includegraphics[width=0.8\linewidth]{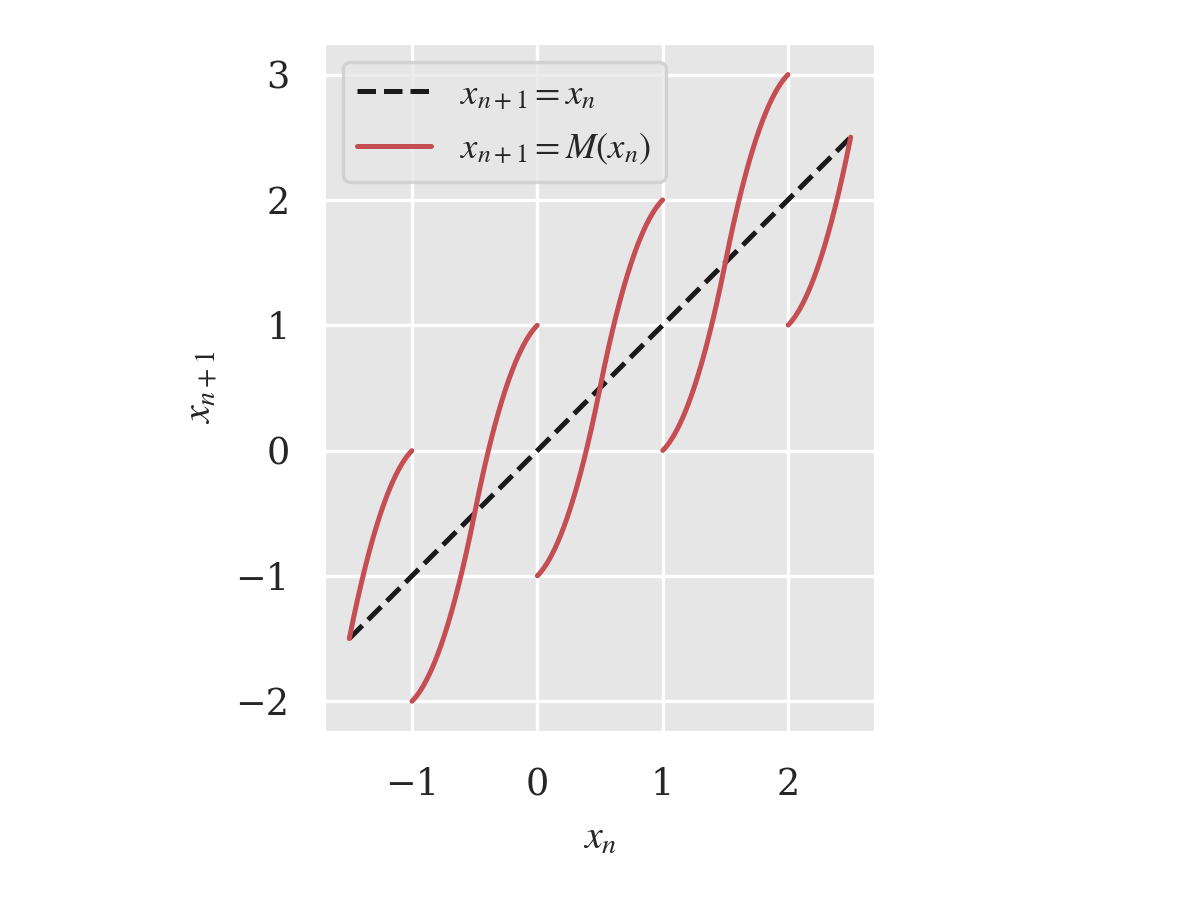}
\caption{The superdiffusive extension of the Pomeau-Manneville map for
  the case $z=2$, {$R=1$} (solid, red online). In black (dashed)
  is also shown the line
  $x_{n+1}=x_n$ for comparison.}
\label{fig:extendmap}
\end{figure}

The {diffusive} {extension} of the map we consider
is {given by}
\begin{equation} \label{eq:pm}
  M(x) = x + R((2x)^z - 1) \quad \text{for} \quad 0<x<\frac{1}{2}\quad,
\end{equation}
{lifted} across the real line by {degree one}
\begin{equation}
M(-x)=-M(x) \quad \textrm{ and } \quad M(x+1)=M(x)+1.
\end{equation}
{This version of the map was studied extensively in
  \cite{schulz_parameter-dependent_2020}, and is closely related to
  systems investigated in \cite{pegler_anomalous_2017,
    zumofen_scale-invariant_1993,ZuKl93b,WaHu93,ACL93}.}  {The
  implicit variable prefactor of $a=R2^z$ ensures that $M$ maps
  $[0,\frac12]$ onto $[-R,\frac12]$, regardless of $z$. Therefore, $R$
  defines the `height', or the distance outside of the unit box that
  the map extends} \cite{dcrc}, which in turn directly parameterizes
the size of the `coupling regions' between different boxes (see
Fig.~\ref{fig:extendmap}).
At $z=1$, the map reduces to a piecewise linear map, the dynamics of
which are well understood, eg.\ \cite{klages_deterministic_1995,
  klages_microscopic_2007,klages_simple_1999,klages_simple_1995,dcrc}. Note
that this definition of the map moves the function $M(x)$ at integer
values, where the original PM map mod $1$ has marginal fixed points,
outside of the boxes into the coupling regions, which provides a
mechanism to potentially generate superdiffusion. Keeping these fixed
points outside of the coupling regions within the boxes instead
generates subdiffusion \cite{GeTo84,geisel_accelerated_1985,
  zumofen_scale-invariant_1993}.

\subsection{Numerical methods\label{sec:meth}}

Numerical computation of the MSD $\langle x^2 \rangle$ and the GDC
$K(R,z)$ directly from the map is done in two ways: firstly, one may
simply sample the MSD from an ensemble of sample trajectories
\begin{equation}
\mathcal{M}(t) := \frac{1}{N} \sum_{i=1}^{N} \left( x_t^{(i)} - x_0^{(i)} \right)^2
\end{equation}
taking each $x_0^{(i)}$ randomly according to some distribution
on the unit interval. Then since we anticipate
$\mathcal{M}(t)\sim t^\beta$, we estimate
\begin{equation} \label{eq:q}
\hat{\beta} = \lim_{i\to\infty} \frac{\log \mathcal{M}(t_{i+1}) - \log \mathcal{M}(t_i)}{\log (t_{i+1}) - \log t_i}
\end{equation}
(for a sequence of sampling times $t_i$ to be selected appropriately -- here
we choose them logarithmically so that the denominator in \eqref{eq:q}
is constant), from which follows
\begin{equation}
\hat{K}(R,z) = \lim_{t\to\infty} \frac{\mathcal{M}(t)}{t^{\hat{\beta}}}.
\end{equation}
This method was used extensively in
\cite{schulz_parameter-dependent_2020} (which contains further details
of the method) and for much of Sec.~\ref{sec:newsec} in which we study
the dependence of these quantities on $R$.  For the variation of $z$,
in Sec.~\ref{sec:results}, {we use} the Taylor-Green-Kubo (TGK)
formula
\cite{kubo_fluctuation-dissipation_1966,dorfman_introduction_1999}
which expresses the MSD in terms of an exact summation formula over
correlations between `velocities'
\cite{korabel_fractal_2007,klages_deterministic_1995,
  KnKl11b,klages_understanding_2002} that, in a time-discrete setting,
we take as displacements per timestep. We found this approach to be
more computationally efficient than other approaches, and a particular
advantage is that this summation can also be broken down into its
constituent parts to analyse the effects of higher-order correlations,
as we will do in Secs.~\ref{sec:results} and \ref{sec:disc}.  The
method works as follows: by defining
\[ v_k := x_{k+1} - x_k \implies x_t - x_0 = \sum_{k=0}^{t-1} v_k \]
we produce the expansion
\cite{kubo_fluctuation-dissipation_1966,dorfman_introduction_1999}
\begin{align} \label{eq:tgk}
  K(z) &= \lim_{t\to\infty} \frac{1}{t^\beta} \langle (x_t - x_0)^2 \rangle \nonumber \\
  &= \lim_{t\to\infty} \frac{1}{t^\beta} \sum_{k=0}^{t-1} \sum_{l=0}^{t-1} \langle v_k v_l \rangle \nonumber \\
  &= \lim_{t\to\infty} \frac{1}{t^\beta} \left[ \sum_{k=0}^{t-1} \langle v_k^2 \rangle + 2 \sum_{k=0}^{t-1} \sum_{j=1}^{t-k-1} \langle v_k v_{k+j} \rangle \right].
\end{align}
{Care must be taken in simplifying this expansion any
further, as in our case, unlike in many systems,
$\langle v_k v_{k+j} \rangle$ is not necessarily stationary.}
As is typical \cite{klages_deterministic_1995,klages_understanding_2002},
we consider only the integer part of each $x_k$, and neglect
the fractional part. This choice eliminates microscopic
(within each unit `box') correlations and ensures that
the first term can be understood as a simple random walk
result that, for the present setting, correctly reproduces
the exact value of $K(1)=\frac23$, thus giving a physically
meaningful `lowest order' approximation
\cite{klages_deterministic_1995,dcrc,klages_understanding_2002}.
Naturally, for $t\to\infty$ the same asymptotic result
is given by either approach.

{In both methods,} {the MSD and GDC are
dependent on} {taking} {an ensemble average,
the outcome of which {may depend}
strongly on the distribution of initial conditions} {used}
{in the ensemble (see Sec.~\ref{sec:aging} for details).}
For our numerical results
{in Secs.~\ref{sec:newsec} and \ref{sec:results}}
we average over an ensemble of simulated trajectories
{(at least $N=10^4$),} each of which is initialised
with $x_0$ {distributed uniformly} over the unit interval;
{in Sec.~\ref{sec:aging} we consider the case in which
the ensemble is `aged' by some initial time $t_a$, and compare
it to the non-aged case below.
The physical significance of this is that initial conditions
in aged ensembles are in some sense closer to equilibrium
conditions, ie.\ closer to the system's invariant density,
if one exists. In cases where an invariant density does not exist,
the aging time parameter is of interest in its
own right (see Sec.~\ref{sec:aging} for details).}

\subsection{Predictions for LWs from CTRW theory\label{sec:ctrw}}

Anomalous diffusion in this type of deterministic dynamical system
has often been modelled stochastically by CTRWs
\cite{GeTo84,ShlKl85,geisel_accelerated_1985,
  zumofen_scale-invariant_1993,
  Bar03,KCKSG06,korabel_fractal_2007},
a type of random walk in which both the vectors of
displacement, and the time interval between displacement events,
are drawn randomly from a defined joint distribution; see
\cite{klafter_first_2011,metzler_random_2000,
  zaburdaev_levy_2015,KRS08,BoGe90}
for reviews. It is well-studied partly due to the
rich theory of Montroll and Weiss
\cite{montroll_random_1965,montroll_random_1973,scher_anomalous_1975},
which enables many key statistics such as the MSD to be easily extracted.
The classical CTRW approach
\cite{KCKSG06,korabel_fractal_2007,zumofen_scale-invariant_1993,
  ZuKl93b,geisel_accelerated_1985}
models the time spent by a particle in the laminar phase of
motion near the fixed point {of the reduced map} as being
distributed according to the
power law (Pareto Type II, Lomax, or $q$-exponential
\footnote{under the parameter change $\gamma=\frac{2-q}{q-1}$,
for $1<q<2$ \cite{picoli_jr_q-distributions_2009}} distribution)
\begin{equation} \label{eq:wt}
w(t) = \frac{\gamma b^\gamma}{(b+t)^{1+\gamma}}
\end{equation}
for $\gamma := 1/(z-1)$ and $b := \gamma/a$. This distribution is
generated from the map by the following simplified reasoning
\cite{GeTo84,geisel_accelerated_1985,zumofen_scale-invariant_1993,korabel_fractal_2007}:
The motion of a particle in the neighbourhood of the fixed point may
be modelled as a continuous laminar motion, obeying a differential
equation
\[ \dv{x}{t} \approx x_{n+1} - x_n = ax^z \]
which is easily solved. The time taken to `escape' from the fixed
point may be determined from this motion as the time taken to reach
$x(t)=1$ from a given injection point $x(0)=x_0$
\cite{korabel_fractal_2007}, and the distribution of escape times
follows by assuming a uniform injection density of $x_0$.

In the CTRW model (here taken in one dimension only),
a particle is initialised at the origin at time $t=0$,
and, for a sequence of time durations $t_i$ drawn i.i.d.\ from
the distribution with density $w(t)$, moves with a
constant velocity in a random direction (left or right,
with equal probability) for a duration $t_i$.
At the points between time intervals $t_i$, a new direction
is chosen independently. {In this sense, the CTRW
defines a `semi-Markov' process, in which the displacements
at each step are Markov with respect to the step number,
but not with respect to the physical system time.}
Details of how this method is applied to our case
are given in Appendix~A, and for further details of this
method see references
\cite{zumofen_scale-invariant_1993,korabel_fractal_2007,
  pegler_anomalous_2017,klafter_first_2011,
  zaburdaev_levy_2015,ZuKl93b}.


Using the CTRW defined above gives the MSD of what is called a LW
\cite{SKW82,ShlKl85,KBS87,Shles87,zumofen_scale-invariant_1993,ZuKl93b}
to leading order as
\begin{equation*} 
  \langle x^2 \rangle \sim v_0^2
  \begin{cases}
    \frac{2b}{\gamma-2}\,t, & 1<z<\frac{3}{2}, \\ 
    \frac{2b^{\gamma-1}(\gamma-1)}{(3-\gamma)(2-\gamma)}\,t^{3-\gamma}, & \frac{3}{2}<z<2, \\
      (1-\gamma)\,t^2, & z>2. 
  \end{cases}
\end{equation*}
The time-dependence on $t^\beta$ is very well known
\cite{geisel_accelerated_1985,ACL93,zumofen_scale-invariant_1993,
  ZuKl93b,WaHu93,DeCvi97,BaKl97,DeDa98,Bar03,KCKSG06,korabel_fractal_2007}.
The multiplicative constant yielding the GDC
was first reported in \cite{ZuKl93b}, more recently
within the context of higher-dimensional LWs in the
suppemental material of \cite{zaburdaev_superdiffusive_2016},
and in the context of generalised LWs with non-constant
velocity in \cite{albers_exact_2018,albers_nonergodicity_2022};
{we discuss this quantity in more detail in Sec.~\ref{sec:aging}.}

We see from the generated formulae that the parameter space of $z$ is
divided into three distinct regimes, {corresponding to three phases
of $\beta(z)$,} at the boundaries of which $K(z)$
{either vanishes or blows up:}
(I) where $1<z<\frac{3}{2}$, for which the diffusion is normal ($\beta=1$);
(II) where $\frac{3}{2}<z<2$, for which superdiffusion is observed
($1<\beta<2$); and (III) where $z>2$,
for which an extreme form of superdiffusion, dominated by ballistic motion,
is observed ($\beta=2$). These emerge from the CTRW calculations as a direct
result of moments of the jump time distribution $w(t)$ ceasing to exist
due to heavy tails: in (I), where $\gamma>2$, $w(t)$ has a well-defined
mean and variance and so the Central Limit Theorem dictates the
diffusion to be normal and with (in the long time limit) a Gaussian
propagator; in (II), $1<\gamma<2$ and therefore $w(t)$ has a first
moment but not a second, resulting in anomalous diffusion; in (III),
$0<\gamma<1$ and therefore not even a first moment exists for $w(t)$,
resulting in ballistic trajectories of unbounded average length.

\section{Variation of coupling length $R$\label{sec:newsec}}

\subsection{On the diffusive exponent $\beta$}

In this section, we review the diffusive behaviour
of the map \eqref{eq:pm} in terms of $\beta$,
under variation of the coupling length $R$.
We show that, away from the nicely understood special case $R=1$,
the quantity $\beta$ is highly non-trivially dependent on $R$;
we find that superdiffusion is only attained for some very specific
values, and diffusion is completely suppressed in others.
For these purposes we fix $z\in\{1.25, 1.75, 2.5\}$,
thus exploring typical behaviours in each of the map's
three main dynamical regimes.

Results from numerical simulations for the quantity
$\hat{\beta}$ are shown in Fig.~\ref{fig:schulz_beta}.
From this we draw the following conclusions:

\begin{figure*}[t]
\centering
\includegraphics[width=0.32\linewidth]{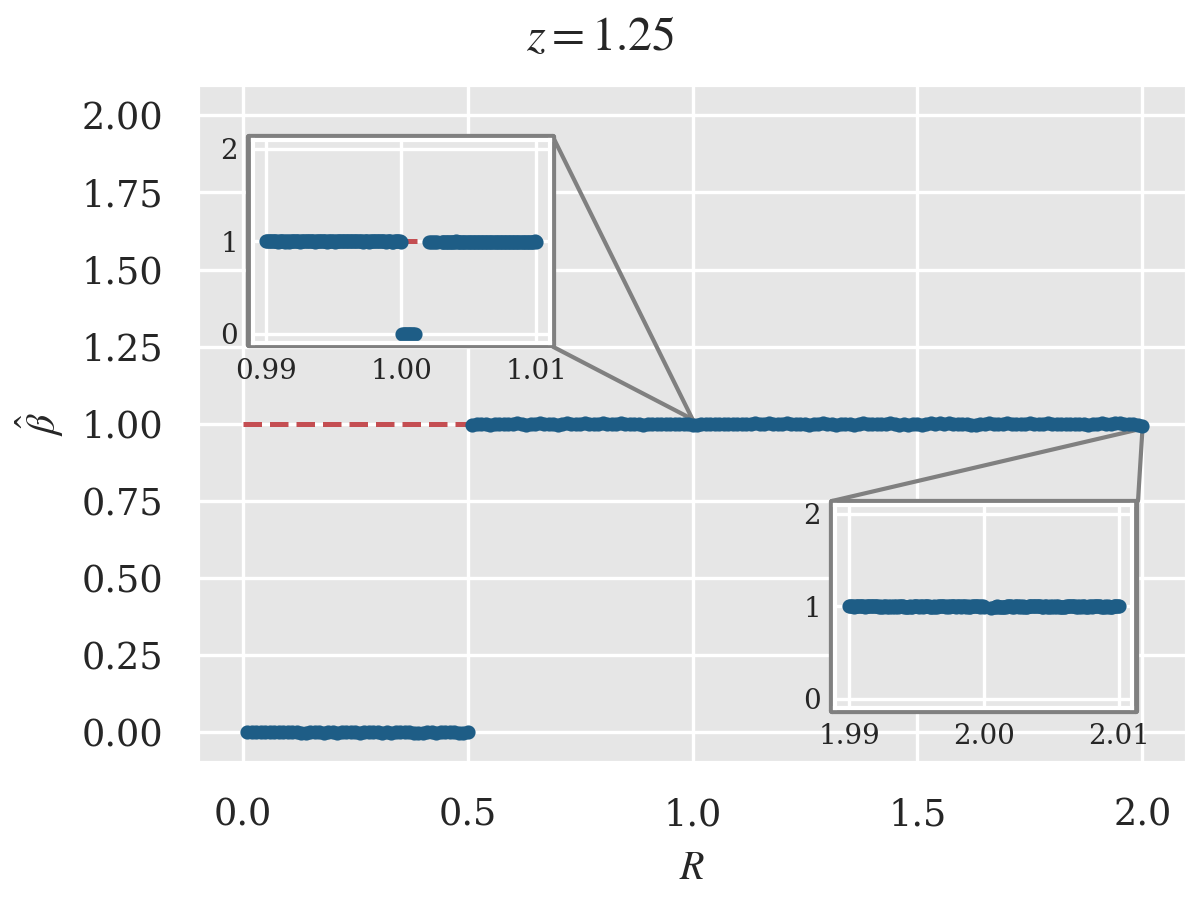}
\includegraphics[width=0.32\linewidth]{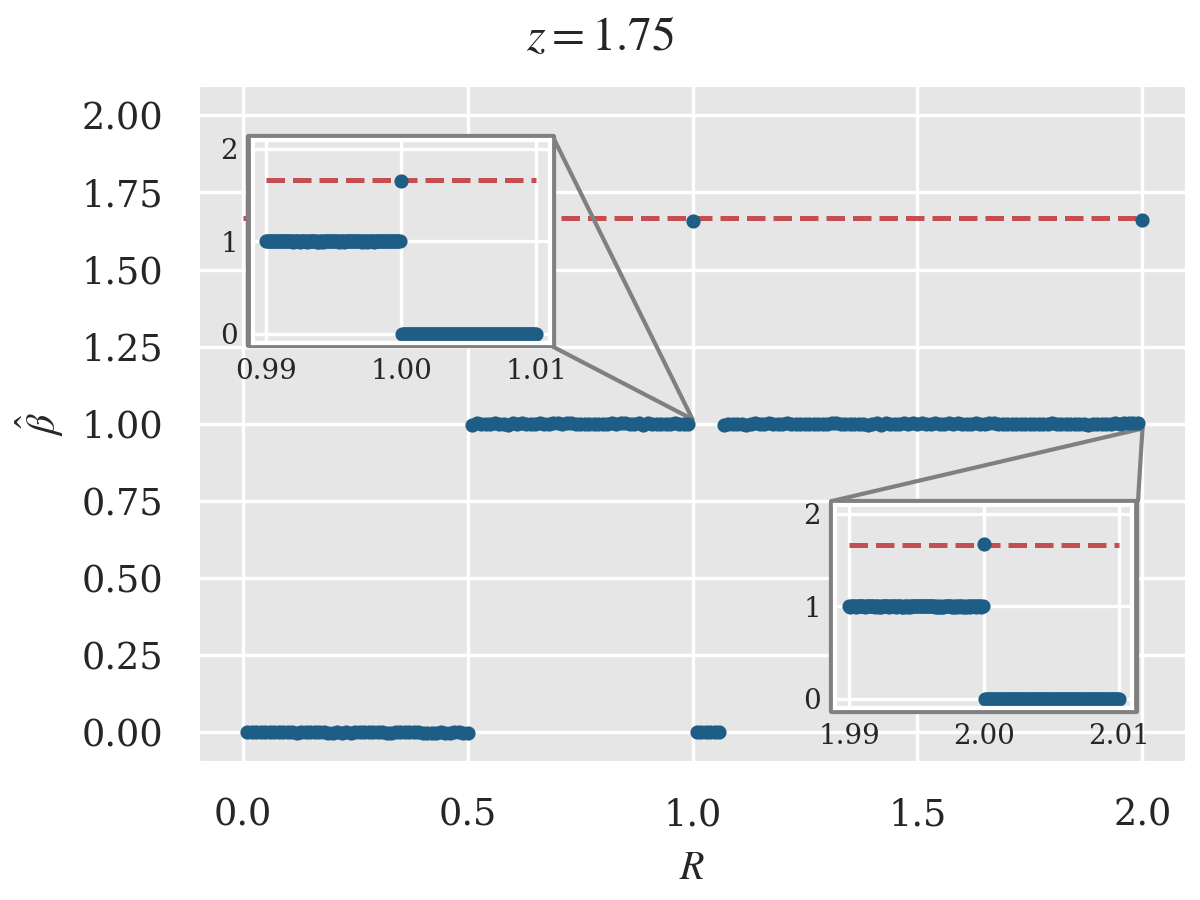}
\includegraphics[width=0.32\linewidth]{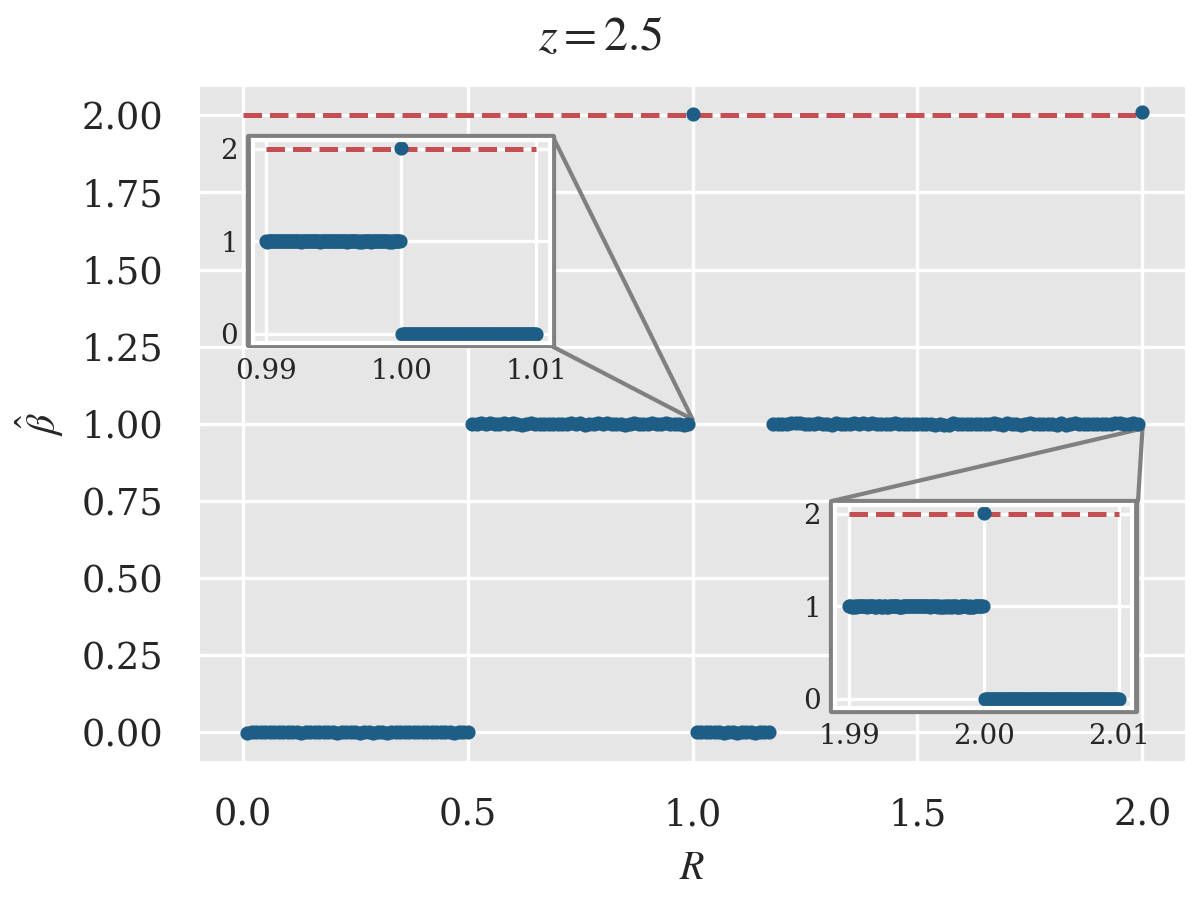}
\caption{Numerically computed $\hat{\beta}(R,z)$ for
  $z\in\{1.25,1.75,2.5\}$ (from top), computed from an ensemble of
  $N=10^5$ initially uniformly distributed sample trajectories aged
  by $T_a=10^6$ timesteps. Main figures are computed for 201 values
  of $R\in[0,2]$ ($\Delta R=0.01$); insets are computed with 201 values
  each of $R\in[0.99,1.01]$ and $R\in[1.99,2.01]$ respectively
  ($\Delta R=10^{-4}$). Dashed lines (in red) indicate predictions
  given by CTRW theory in \eqref{eq:results}.}
\label{fig:schulz_beta}
\end{figure*}

For all $z$, if $R\leq\frac12$, then trivially $\beta=0$,
since all unit boxes $[k-\frac12, k+\frac12]$, $k\in\Z$
are positively invariant, and therefore the dynamics are completely
localised. For $z=1$, for all other $R>\frac12$, the dynamics are
normally diffusive ($\beta=1$; cf.\ \cite{klages_deterministic_1995,
klages_microscopic_2007,klages_simple_1999,klages_simple_1995,dcrc}).

For $z>1$, if $R\in\N$, then we attain the superdiffusion
described in Sec.~\ref{sec:ctrw}. This is possible because in this case,
the map's marginal point maps onto a copy of itself, displaced by $R$ steps
to the left or to the right.
This allows for extended `runs' in one direction or the other,
generating superdiffusion \cite{zumofen_scale-invariant_1993,
  geisel_accelerated_1985}.  For \emph{most} other values $R\notin\N$,
however, the diffusion is normal; since if the marginal point of $M$
does not map onto itself, all higher iterates of the map $M^2, M^3,
\dots$ are uniformly hyperbolic.

\begin{figure}
\centering
\includegraphics[width=0.8\linewidth]{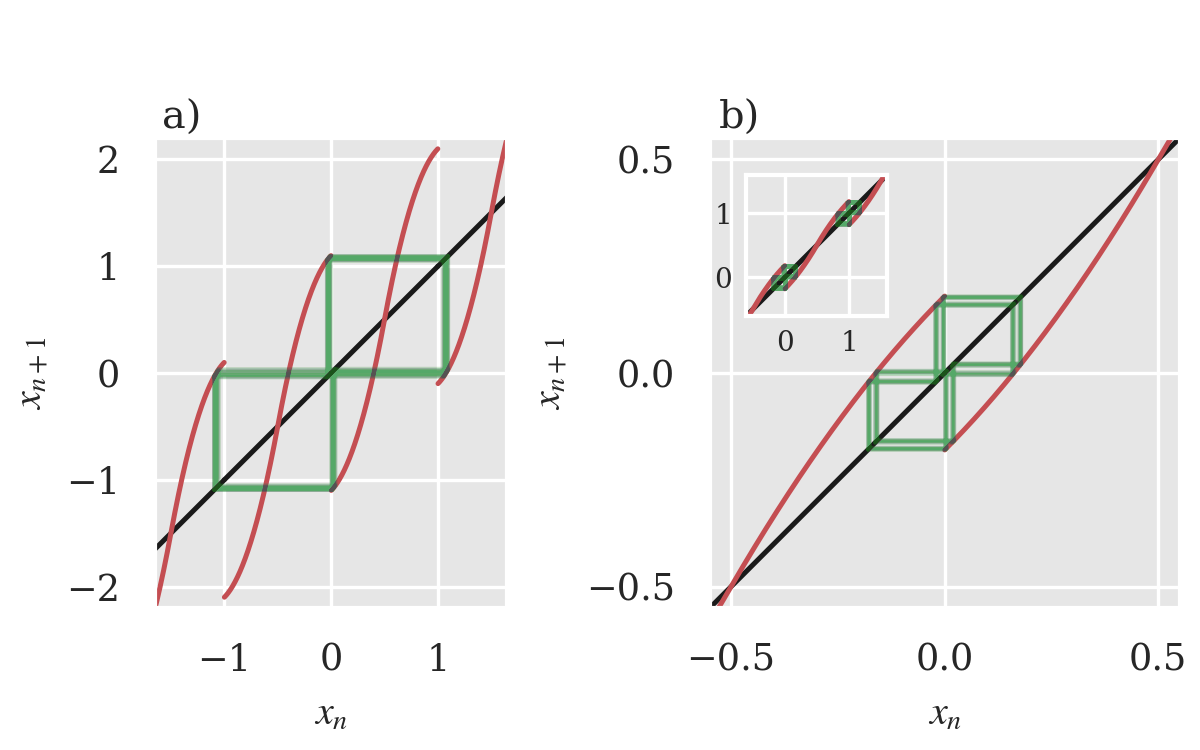}
\caption{Cobweb plots showing two examples of special invariant sets.
  (a) The region $[-x_c, x_c]$ is invariant under $M^2$ in the
  map with $z=2$, $R=1.1$, causing $\beta=0$. (b) Orbits in the map
  with $z=2$, $R=0.18$ localise around four disconnected intervals,
  close to $x\in\Z$, producing the bifurcation in
  Fig.~\ref{fig:schulz_main}.}
\label{fig:schulz_special}
\end{figure}

However, for $z>1$, there is not always normal diffusion
when $R\notin\N$: we observe intervals of the form
$R\in (\ell, \ell+\delta_c]$, $\ell\in\N$, $\delta_c = \delta_c(R,z)$
in which $\beta$ completely vanishes.
This is because if $x_0$ is in a neighbourhood of the origin,
then $M(x_0)$, rather than mapping onto a neighbouring branch pointing
in the same direction, as in superdiffusion, instead maps onto the
reverse branch, so that $M^2(x_0)$ is returned once again to the origin,
see Fig.~\ref{fig:schulz_special}(a).
Under these conditions, it is possible for there to exist some $x_c(R)$
such that $M^2([-x_c,x_c]) \subseteq [-x_c,x_c]$, therefore causing
the dynamics to remain localised. Finally, 
since the remainder of the map is repelling,
almost all trajectories are eventually absorbed into this localisation cycle.
On the other hand, if $R>k+\delta_c$, such a $x_c$ may not exist,
and normal diffusion is returned.

Let us denote the ranges of $R$ where diffusion is suppressed
in this way by
\begin{equation} \label{eq:nkz}
\mathcal{N}_k(z) := \{ R\in (k,k+1] : \beta(R,z)=0 \}, \quad k\in\N.
\end{equation}
Trivially $\mathcal{N}_0(z)=(0,\frac12]$ as remarked above. However,
we also now see that for $z>1$ and $k\geq 1$, zero-sets exist
of the form $N_k(z)=(k, k+\delta_c]$.
Based on our limited numerical evidence,
we suggest the following claims: if $\mu$ is the Lebesgue measure,
\begin{enumerate}
\item $\mu(\mathcal{N}_{k+1}(z)) < \mu(\mathcal{N}_k(z))$
(zero-sets get smaller as we increase $k$);
\item $\mu(\mathcal{N}_k(z)) < \mu(\mathcal{N}_k(z'))$ if $z<z'$
(zero-sets get larger as we increase $z$).
\end{enumerate}
As $z$ increases, the fixed point
becomes `stickier', thus widening the
range of $R$ for which
$x_c$ can exist;
conversely, for larger $k$, the expanding
gradient of the map is larger,
reducing the marginal point's stickiness.

In conclusion, we see that, far from the idealised picture presented
by the well-studied $R=1$ scenario, there is instead a wide
range of diffusive behaviours that can be observed,
including normal diffusion and localisation,
of which superdiffusion is a rare special case.  Further, we note that
$\beta$ is highly discontinuous in $R$, meaning that superdiffusion is
not dynamically stable to small perturbations of the map
(cf.\ \cite{knight_capturing_2011}). One may now wonder about the
functional form of the associated GDC under variation of $R$ and $z$,
which has been studied numerically in
\cite{schulz_parameter-dependent_2020}. As is already well-known from
previous works on parameter-dependent diffusion in simple
one-dimensional maps, whenever diffusion is normal for the PM map the
GDC appears to be a fractal function of $R$; see
\cite{klages_microscopic_2007} for a review and
\cite{klages_simple_1995,klages_simple_1999,dcrc} for the special case
$z=1$. Whenever there are localisation regions with $\hat{\beta}=0$
the GDC jumps to zero while superdiffusive parameter values of $R$
correspond to local maxima in the GDC. How precisely these transitions
happen in terms of continuity properties of the GDC as a function of
$R$ remains an interesting open question, as this is very difficult to
resolve in computer simulations. Similar transitions to localisation
regions related to bifurcation scenarios have been reported in
\cite{korabel_fractality_2004,korabel_fractal_2002} for a nonlinear
and in \cite{GrKl02} for a biased piecewise linear map.

\subsection{Bifurcation scenario}

\begin{figure*}
\centering
\includegraphics[width=0.32\linewidth]{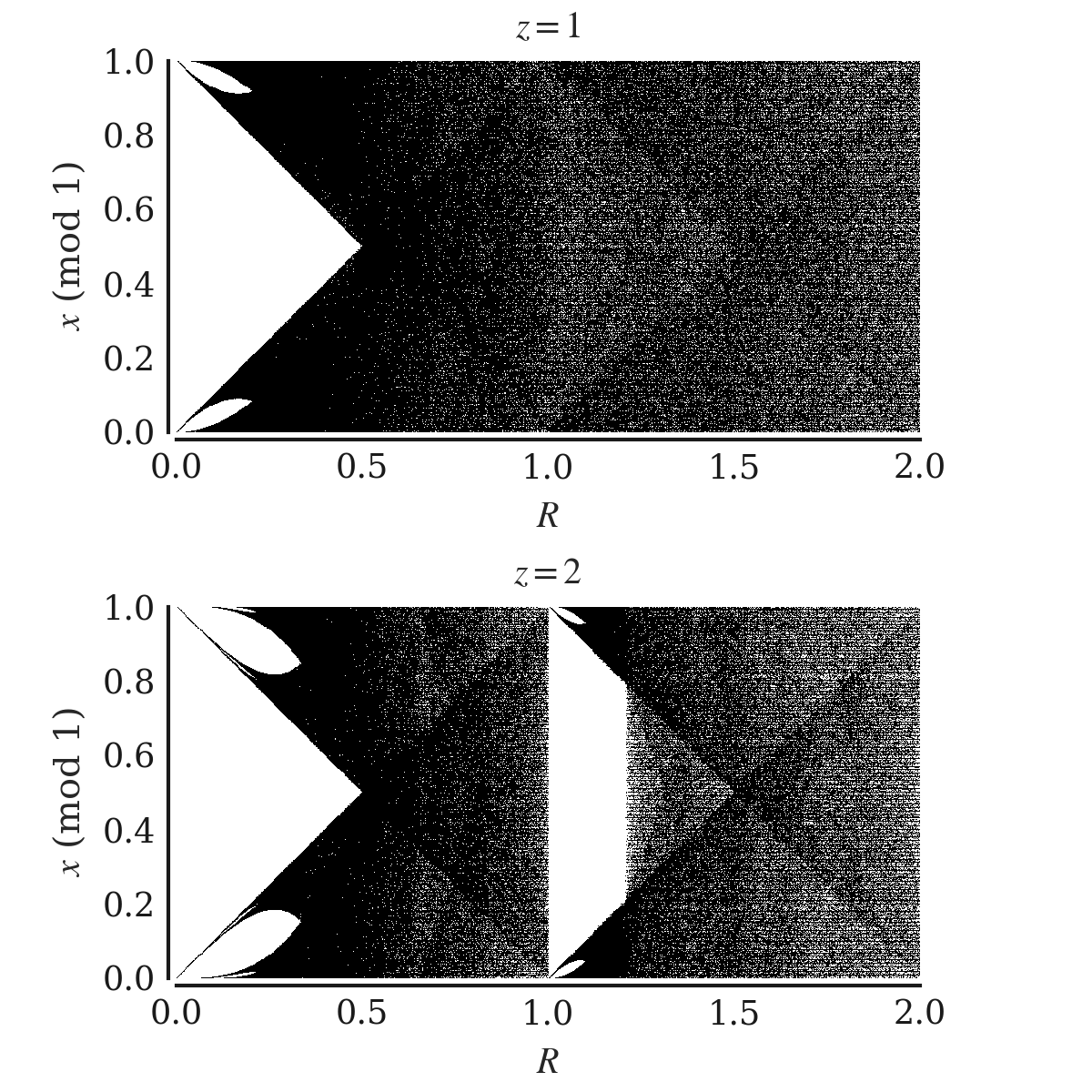}
\includegraphics[width=0.32\linewidth]{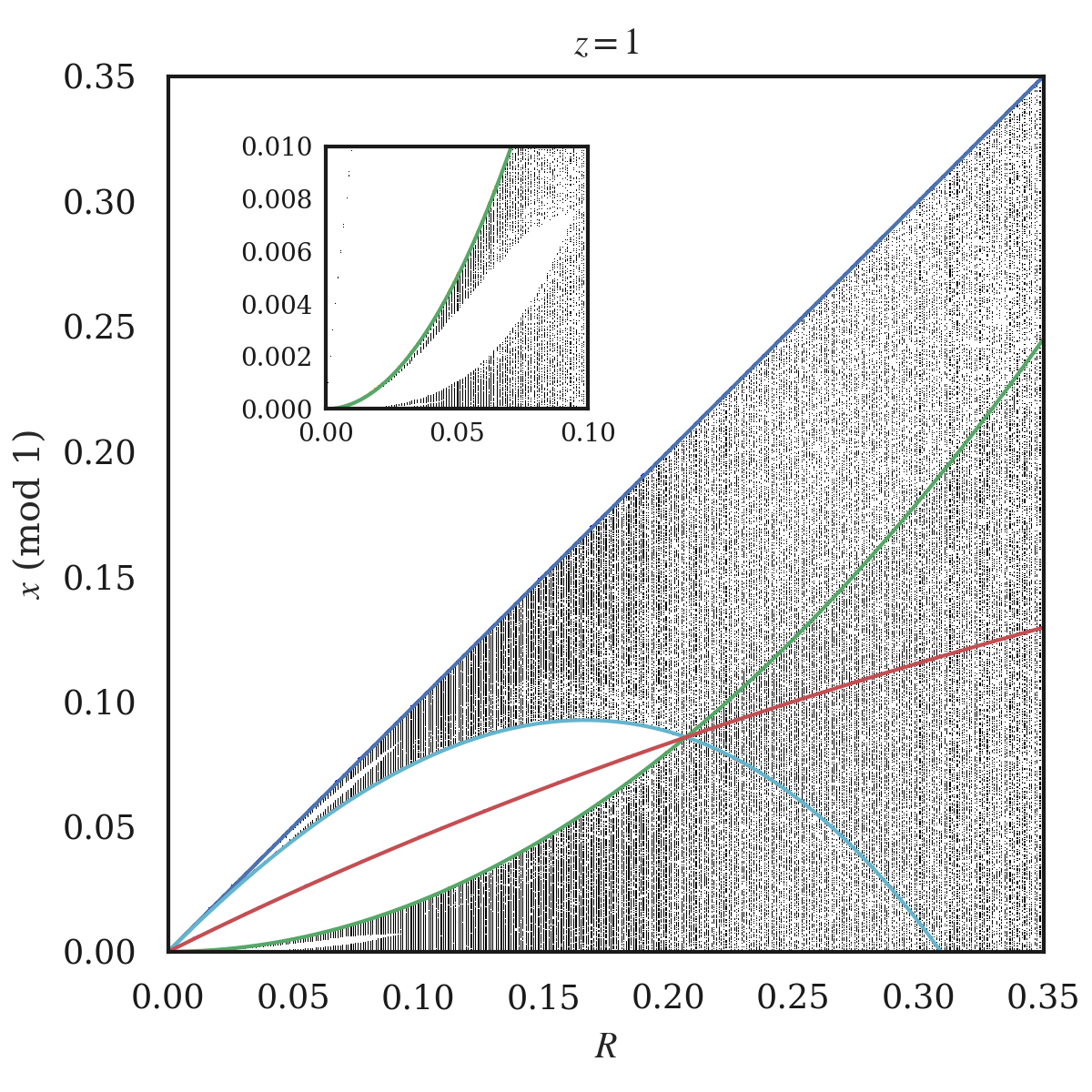}
\includegraphics[width=0.32\linewidth]{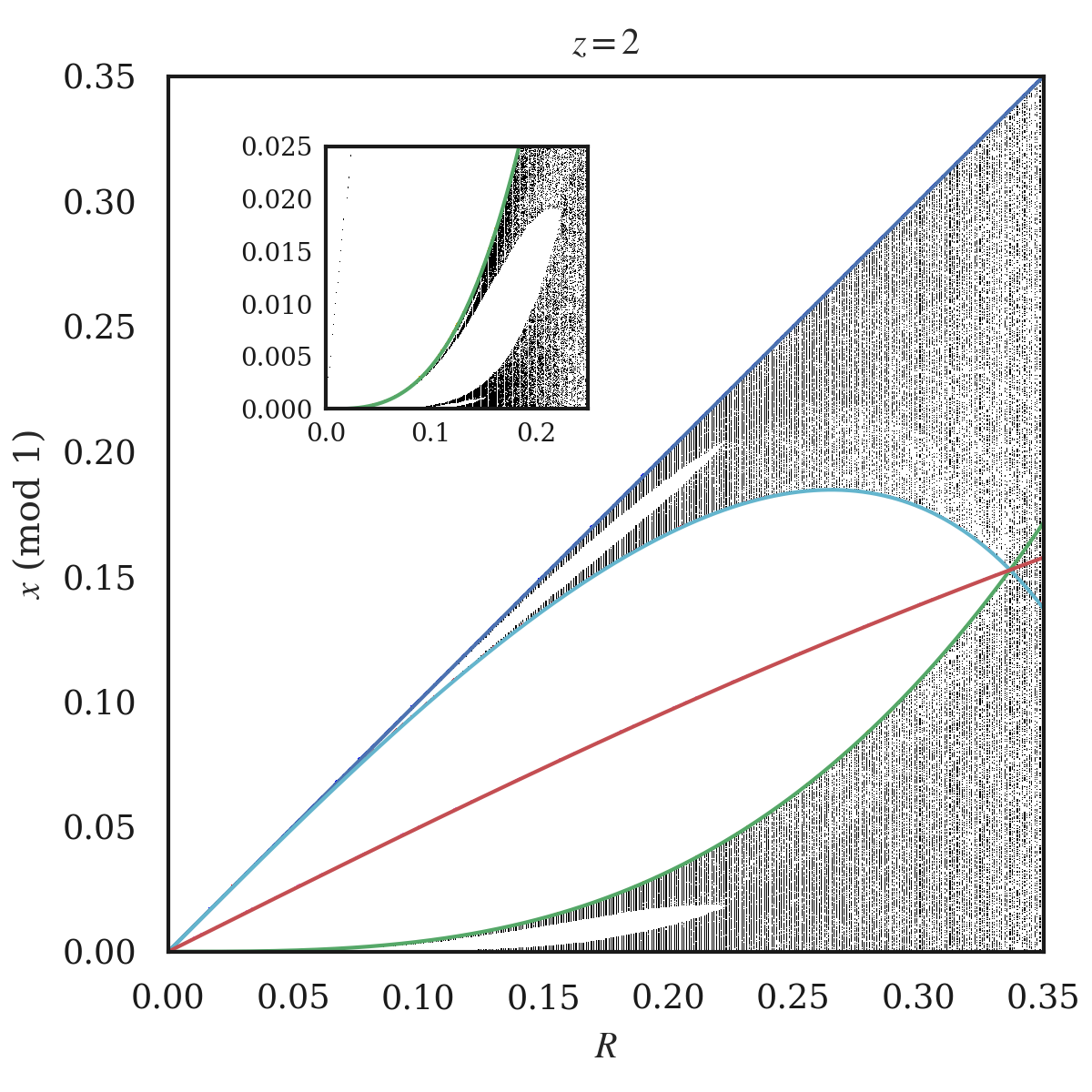}
\caption{Left: Orbits {of the map $M(x)$ (mod 1) \eqref{eq:pm},
  for} $N=10^3$ sample trajectories running for
  $T=10^7$ timesteps, for $2\times 10^3$ values of $R\in[0,2]$
  ($\Delta R=10^{-3}$), for $z\in\{1,2\}$ (from top). Observe the
  complex bifurcation scenario that is present for small $R$,
  and the qualitative difference observed under change of $z$.
  Center and right: Zoomed portions of left, in comparison
  with the dynamic contours {mentioned in the text, and
  calculated in the appendix:} the boundary of $M$
  given by $R$ in blue; the boundaries of $M^2$, namely $x_\alpha$ and
  $x_\beta$, in green and cyan respectively; and the unstable fixed
  point $x_u$ of $M^2$ in red. The contours of $M^2$ collide at the critical
  point $R_c$, see text. Insets show a further zoomed portion indicating
  an iterative hierarchy of bubble-like structures.}
\label{fig:schulz_main}
\end{figure*}

Further, in this section we explore a complex bifurcation scenario
that the map undergoes under transition of $R\in \mathcal{N}_k(z)$. In this
region, the map's dynamics are localised, causing $\beta=0$; we examine
this localisation to reveal a detailed fractal bifurcation structure.
In order to assess this in the nonlinear PM map, we
first examine it as it manifests in the linear map when $z=1$.

In Fig.~\ref{fig:schulz_main} are shown stable orbits, modulo 1,
  of the Pomeau-Manneville map $M(R,z)$ under continuous variation of
  $R$, for $z\in\{1,2\}$. (These values of $z$ are chosen for ease of
  calculation.) In cases where we have normal diffusion we have
  (strong) chaos (in the sense of positive Lyapunov exponents),
and these orbits cover the entire unit interval.  However, as noted
earlier, when $R\in \mathcal{N}_k(z)$, orbits are localised to
particular subregions of the state space.  For $z=1$, the stable
regions are, by trivial inspection of the map (see
Fig.~\ref{fig:schulz_special}(b) inset, for example) $\bigcup_{j\in\Z}
[j-R, j+R]$.  At $x=j+\frac12$, $j\in\Z$, the map has a repelling
(unstable) fixed point.  At $R=\frac12$, the boundaries of the stable
sets connect with each other \emph{and} coincide with the unstable
fixed point, at which point {normal diffusion} is achieved.
{In bifurcation theory, this marks a crisis point
  \cite{ott_chaos_1993}.}  As discussed above, for $z>1$ we see a
repetition of this bifurcation when $R=k+\delta$, $k\in\N$,
as $R$ passes through $\mathcal{N}_k(z)$, after which
{normal diffusion} is abruptly restored at some point $R_d$.
The exact point at which this occurs is calculated analytically
in Appendix~B, for the case $z=2$,
and we believe this may coincide with the end of $\mathcal{N}_k(z)$; ie.,
normal diffusion is restored once all areas of phase space, in particular
the repelling area, are reachable, see Fig.~\ref{fig:rc} in the Appendix.

However, we see that this is not the full bifurcation picture.
Analysing the map's
second iterate $M^2$, we find another
bubble-like region
in the interval $[0,R]$, and symmetrically in
$[-R,0]$ (appearing in Fig.~\ref{fig:schulz_main} as $[1-R,1]$),
dividing the stable orbits into two smaller disjointed intervals.
As can be seen from the map, this is due to two further unstable fixed
points of $M^2$ (unstable periodic points of $M$) appearing in these regions.
At some critical value $R_c$,
the boundaries of these regions coincide, and {the two regions merge.}
This separation manifests as a 'bubble' in
the left-most corners of the plots in Fig.~\ref{fig:schulz_main}; the centre
and right plots in that figure show close-up images of these regions.

In the cases $z=1$ and $z=2$ we are able to calculate exactly
the dynamical boundaries of these stable regions,
{which we denote by $x_{\alpha}(R)$ and $x_{\beta}(R)$,
see Fig.~\ref{fig:schulz_main}.}
In Appendix~B, we calculate exact values of $x_\alpha(R)$ and
$x_\beta(R)$, and the unstable periodic point $x_u(R)$, analytically,
and deduce the point $R_c$ at which it is shown they coincide.  These
three critical quantities are shown as coloured curves on
Fig.~\ref{fig:schulz_main}. Further, we see in these figures (see
insets) that these subregions divide further for smaller $R$, creating
another hierarchy of bubbles, the boundaries of which are determined
by the third iterate $M^3$ of the map.  (In
Fig.~\ref{fig:schulz_special}(b) this third-order bubble can be seen
quite clearly separating the stable orbits shown.) We conjecture that
this hierarchy of bubbles is infinite, producing a fractal bifurcation
structure; in Fig.~\ref{fig:schulz_large} in the Appendix we show a
series of zoomed-in plots for a range of values of $z$ depicting
increasingly higher orders of this structure.  This result should be
viewed in the light of known bifurcation structures in other, similar
diffusive maps, such as the climbing sine \cite{korabel_fractal_2002,
  korabel_fractality_2004}. and a biased piecewise linear map
\cite{GrKl02}.

\section{Variation of nonlinearity $z$\label{sec:results}}

\subsection{Results}

\begin{figure}[t]
\centering
\includegraphics[width=0.7\linewidth]{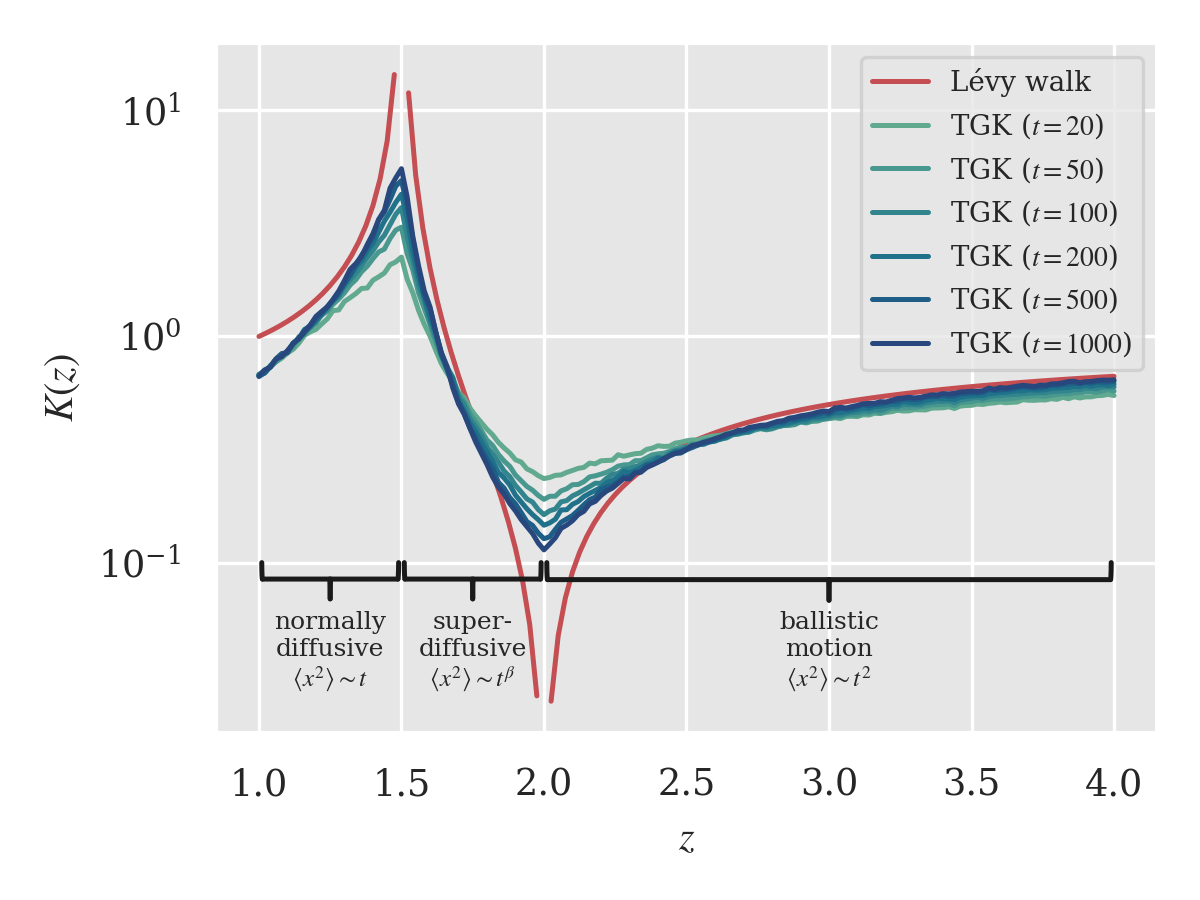}
\caption{The GDC $K(z)$ \eqref{eq:gdc} generated from TGK expansions
  \eqref{eq:tgk} of the superdiffusive PM map \eqref{eq:pm}, { fixing $R=1$,} using
  increasingly many terms $t$ of the expansion and the analytic
  estimate for $\beta$ obtained from \eqref{eq:results} (rough curves,
  $t$ increasing from light to dark). Also shown is
  the LW approximation taken from \eqref{eq:results} and \eqref{eq:gdc}
  (smooth curve, red online). {The different dynamical regimes shown in
  \eqref{eq:results} are annotated on the plotting area.}
  All plots are generated from an ensemble of
  $10^4$ simulated trajectories and consist of $151$ points.}
\label{fig:tgkconvfresh}
\end{figure}

Using the CTRW defined {in Sec.~\ref{sec:ctrw}} gives the MSD of
the LW to leading order as $a=2^z$,
\cite{zumofen_scale-invariant_1993,ZuKl93b,WaHu93,ACL93,
  geisel_accelerated_1985,ShlKl85,TG04}
\begin{equation} \label{eq:results}
  \langle x^2 \rangle \sim v_0^2
  \begin{cases}
    \frac{2b}{\gamma-2}\,t, & 1<z<\frac{3}{2}, \\ 
    \frac{2b^{\gamma-1}(\gamma-1)}{(3-\gamma)(2-\gamma)}\,t^{3-\gamma}, & \frac{3}{2}<z<2, \\
      (1-\gamma)\,t^2, & z>2, 
  \end{cases}
\end{equation}
{for $v_0$ a velocity term; in the case $R=1$, whereby near
the fixed point a particle is displaced one unit per timestep,
we set $v_0=1$; although not commonly studied, via the same
reasoning we may take for other $R\in\N$ the same formulae,
substituting $v_0=R$ appropriately.}
These results are shown in Fig.~\ref{fig:tgkconvfresh},
along with numerical estimates of the same quantity calculated
via the TGK formula in \eqref{eq:tgk}, {as described
in Sec.~\ref{sec:setup}.}
{The numerics are seen to generally fit well to the formulae,
notwithstanding a few remarks which we discuss below:}

At the dynamical transition points $z=\frac32$ and $z=2$,
we observe in the case of the LW that $K(z)$ approaches
infinity and zero respectively. This is because,
considering a more detailed expansion, via the same methodology
\cite{geisel_accelerated_1985, zumofen_scale-invariant_1993},
{at the transition points between dynamical regimes,
logarithmic corrections to the MSD cause the GDC to be
ill-defined at the transition points,}
\begin{equation} \label{eq:log}
  \langle x^2 \rangle \sim \begin{cases} t\log t, & z=\frac{3}{2}, \\
    t^2 / \log t, & z=2, \end{cases}
\end{equation}
which matches to exact results for the PM map obtained by
methods of dynamical system theory \cite{WaHu93,ACL93,TG04}.
Fig.~\ref{fig:tgkconvfresh} shows that around these two
transition points $K(z)$ displays characteristic shapes.
Numerical estimates with finite computation time cannot
overcome the slow logarithmic corrections of \eqref{eq:log}
to reproduce these analytical predictions accurately; but taking
increasingly many terms $t$ from the TGK expansion (which
equates to looking at the process over a longer time)
indeed produces sharper peaks and troughs at the
dynamic transitions (see Fig.~\ref{fig:tgkconvfresh}),
thus confirming {convergence to} the LW scenario
at least qualitatively.

We observe a clear convergence in regime (III) when $z>2$
of the numerical GDC towards the curve generated from the
LW from below (in Sec.~\ref{sec:aging}, we reveal we
also have convergence from above). Likewise, our numerical
study indicates good general matching and a promising
convergence to the LW result in regime (II)
for $\frac32 <z<2$. However, for regime (I) {where} $0<z<\frac32$
we notice a significant discrepancy, which relates to the
limit of $K(z)$ as $z\to 1$. It is
given by the stochastic LW calculations as $K(1)=1$, but is
shown numerically to have the value approximately $\frac23$. Since for
$z=1$ the function $M(x)$ reduces to a linear shift map, the true
value can also be calculated via straightforward analytic means to be
$K(1)=\frac23$, matching to a simple random walk result
\cite{klages_deterministic_1995,klages_simple_1995,dcrc,
  klages_simple_1999, klages_understanding_2002}.
This represents a major problem with the CTRW calculations which, to
the knowledge of the authors, does not appear to be mentioned or
explained in any existing literature, {which instead} typically
treats the CTRW as an accurate representation of the map's dynamics.
{We attempt to give our analysis of this discrepancy below.}

\subsection{Understanding the discrepancy between deterministic dynamics and CTRW theory\label{sec:disc}}

That there should be such a discrepancy is not \emph{per se}
surprising, since it was also shown in the subdiffusive case
that in the regime of normal diffusion, for $z<\frac32$,
the Lévy model is invalid and instead a selectively
constructed random walk was used to match $K(z)$ in this
parameter region \cite{korabel_fractal_2007}. A normally-diffusive
CTRW with exponential waiting times was also applied and found to
match sufficiently well in this regime \cite{korabel_fractal_2007}.
In the superdiffusive case however we were unable to identify any
natural parameters of the system to fit an exponential walk process
to match the numerical results.

The reason for this is as follows: consider the distribution
from which the time durations $t_i$ (described in Sec.~\ref{sec:setup})
are drawn. For the parameters $z<\frac32$ we assume
(based on both numerical and CTRW-derived evidence)
this distribution has at least its first two moments,
ie.\ $\Exp[t_i] = \mu_1 < \infty$ and $\Exp[t_i^2] = \mu_2 < \infty$.
This allows us to write the Laplace transform of the density function
{(cf.\ Sec.~\ref{sec:ctrw}, calculations in Appendix~A, and
\cite{korabel_fractal_2007})} as
\begin{equation} \label{eq:laplace1}
\tilde{w}(s) = 1 - \mu_1 s + \frac{\mu_2}{2} s^2 + \Orb(s^3),
\end{equation}
from which we can derive via the same Montroll-Weiss theory
\begin{equation} \label{eq:laplace2}
\langle x^2 \rangle (t) \simeq \frac{\mu_2}{\mu_1} \, t + o(t),
\end{equation}
implying $K(z) = \mu_2 / \mu_1$. While there are plenty of
distributions which can reproduce the correct limit of
$K(z)=\frac23$ at $z=1$ (eg. an exponential distribution with
{rate} $\lambda=3$),
one would {still} have to motivate from first principles why these
parameters are justified, based on the dynamics of the map.
Further, to obtain the distinct phase transition at $z=\frac32$,
these parameters would also have to undergo a transition which,
again, must be dynamically motivated (eg.,\ we would require
$\lambda\to 0$ as $z\to\frac32$ for an exponential distribution,
{etc.).}

\begin{figure}[t]
\centering
\includegraphics[width=0.7\linewidth]{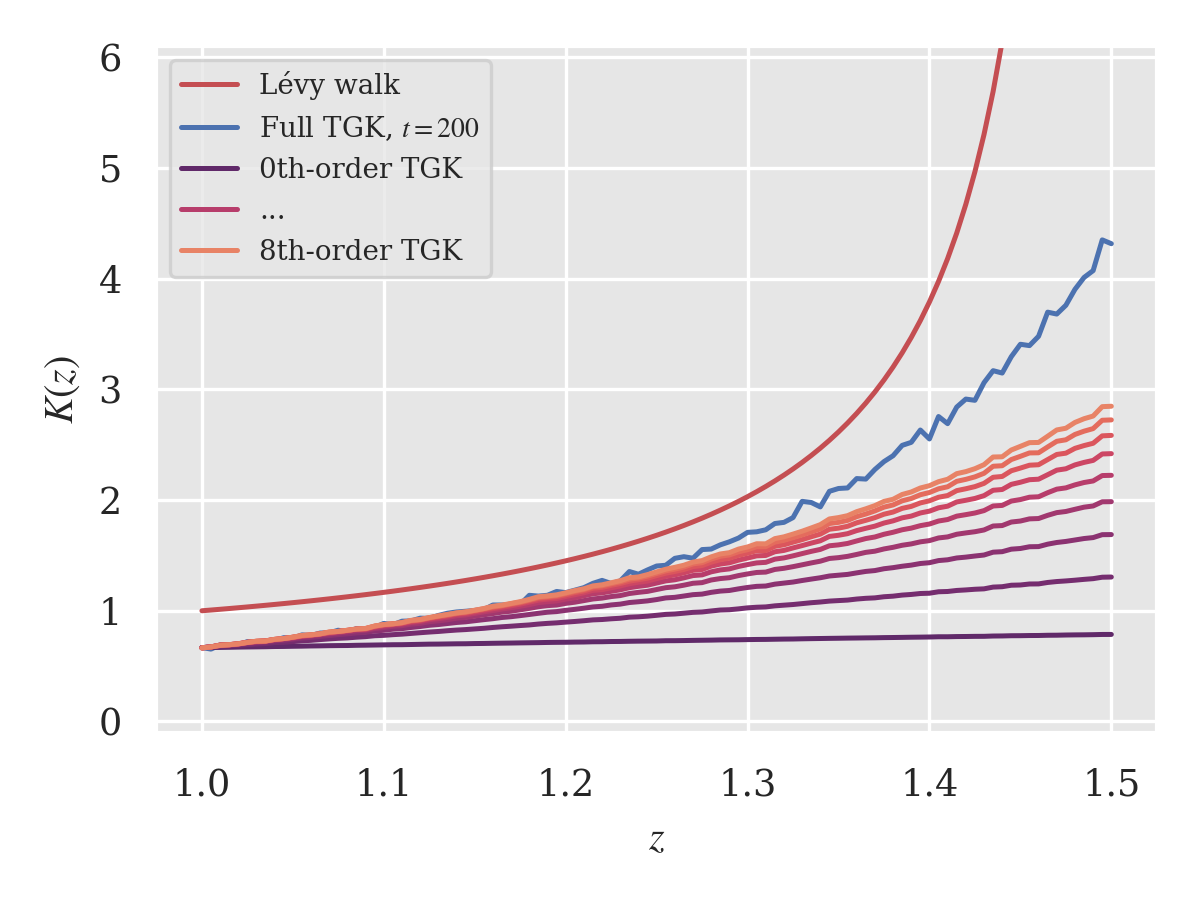}
\caption{$K(z)$ generated from TGK expansions \eqref{eq:tgk} of the
  superdiffusive PM map, { fixing $R=1$,} using increasingly many orders of correlation
  terms $\langle v_k v_{k+j} \rangle$ (lower rough curves, $m$ increasing
  from dark to light). Also shown are the LW approximation
  \eqref{eq:results} (uppermost smooth curve, red online) and the full TGK
  expansion (uppermost rough curve, blue online). All TGK plots are taken
  using $200$ terms of a TGK expansion from an ensemble of $10^4$
  simulated trajectories and consist of $101$ points.}
\label{fig:tgkapprox}
\end{figure}

In general, and in stark contrast to the corresponding subdiffusive
case, we expect that an exponential or Dirac-like {CTRW}
(whose dynamics would generate a simple random walk)
will be unable to reproduce the behaviour of $K(z)$ for $z<\frac32$,
due to high-order correlations which appear in the map's dynamics
for \emph{all} $z>1$, and not only those near the $z=\frac32$ limit.
These are obtained from the decomposition of the TGK {series}
described in Sec.~\ref{sec:setup}. In Fig.~\ref{fig:tgkapprox}
we show this decomposition, with a fixed time parameter of $t=200$,
in the region $1<z<\frac32$, as a series of increasing partial
expansions, starting with the `0th'-order expansion -- comprising
only the first of the two terms in \eqref{eq:tgk} -- and sequentially
adding cross-correlations from the second term of the formula
(we call the `$m$th'-order expansion that in which $j$ is allowed
to range from $1$ to $m$; ie., correlations between displacements
up to $m$ time units apart are considered). This decomposition
reveals the immediate presence of high-order correlations between
timesteps in the map's dynamics, which precludes the modelling of
the process as a random walk.
Only in the $z=1$ limit is the correct GDC obtained if in the TGK
expansion we do not consider correlation terms (in which case, as is
already known, the map can be modelled simply by a random walk).

In principle the incorporation of correlations derived from a
TGK-type expansion into a random walk model could lead to a suitable
representation of the dynamics as a higher-order Markov process,
a correlated random walk, or some other such process. However
the above investigation reveals to us that as $z\to\frac32$, the
required order of correlations under consideration becomes very large.
This indicates to us that the transition between normal diffusion
and superdiffusion in this regime represents a highly non-trivial dynamical
scenario, much more so than its subdiffusive counterpart.
This transition scenario corresponds to the immediate
realisation of the marginal fixed point which is present for all
$z>1$, and causes the map to no longer be hyperbolic.
In \cite{demers_escape_2016} it was shown that for the PM map,
this transition guarantees a rate of escape from the fixed point
which is asymptotically a power law for all $z>1$. In the case $z=1$,
on the other hand, this distribution becomes exponential.

\begin{figure*}[t]
\centering
\includegraphics[width=\textwidth]{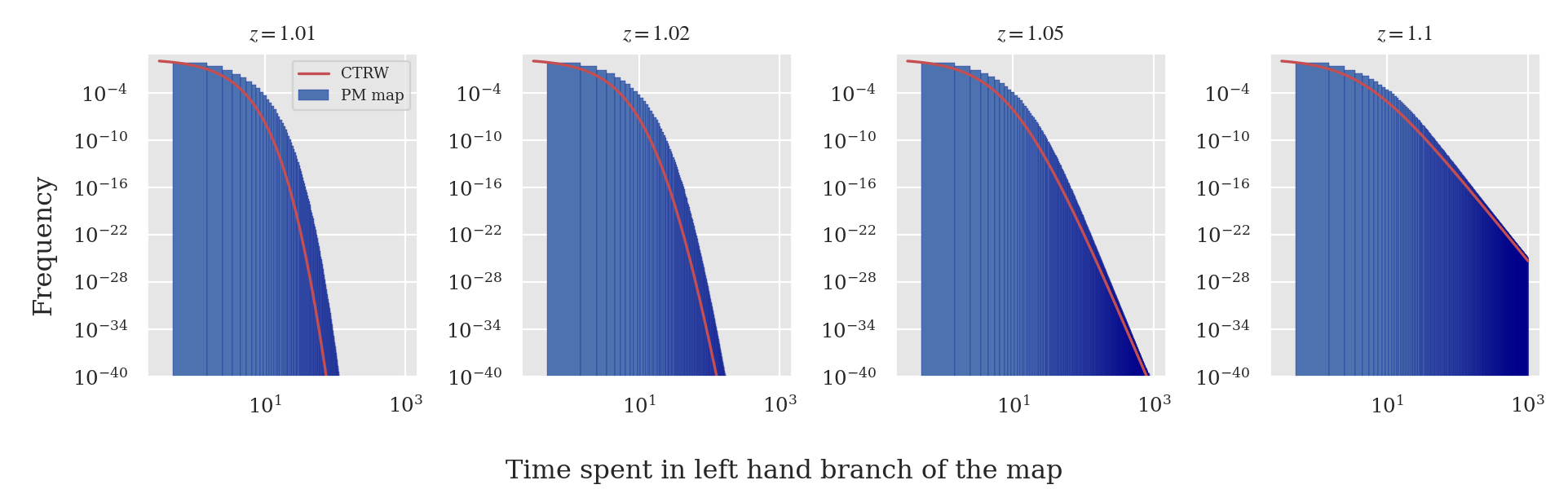}
\caption{Histograms of the time spent in the left-hand branch
  (adjacent to the marginal point) of the PM map, calculated
  from pre-iterates of the map, for varying values of $z$, { fixing $R=1$.} In red for
  comparison is shown the corresponding density $w(t)$ as prescribed
  by \eqref{eq:wt}. 
  Backwards iterates from the map are generated using numerical root finding
  using Python's \texttt{decimal} package, to 40 decimal places.}
\label{fig:hist}
\end{figure*}

We observe this transition scenario in Fig.~\ref{fig:hist}, which shows
the discrepancy between the density $w(t)$, prescribed by CTRW, for the
time between displacement events, and the density of the true number of
timesteps a random point (injected uniformly) on the PM map \eqref{eq:pm}
will take until it leaves the branch containing the marginal fixed point,
determined numerically by iteratively calculating preimages of the escape
point on the marginal branch of the map.
We notice that, for $z>1$, $w(t)$ is indeed asymptotic to the `true'
density, with only a transient difference, but this difference
grows unboundedly in scale and in duration in the limit $z\to 1$,
eventually spanning many orders of magnitude.
Similar transients were noted in relation to power laws in, eg.,
\cite{dahlqvist_escape_1999} and references therein,
and can lead to poor or misleading statistics for a
variety of chaotic systems.
This may also be a contributing factor {to the difference
observed in} the GDC.

\subsection{Effect of aging on the generalised diffusion coefficient\label{sec:aging}}

\begin{figure}[t]
\centering
\includegraphics[width=0.7\linewidth]{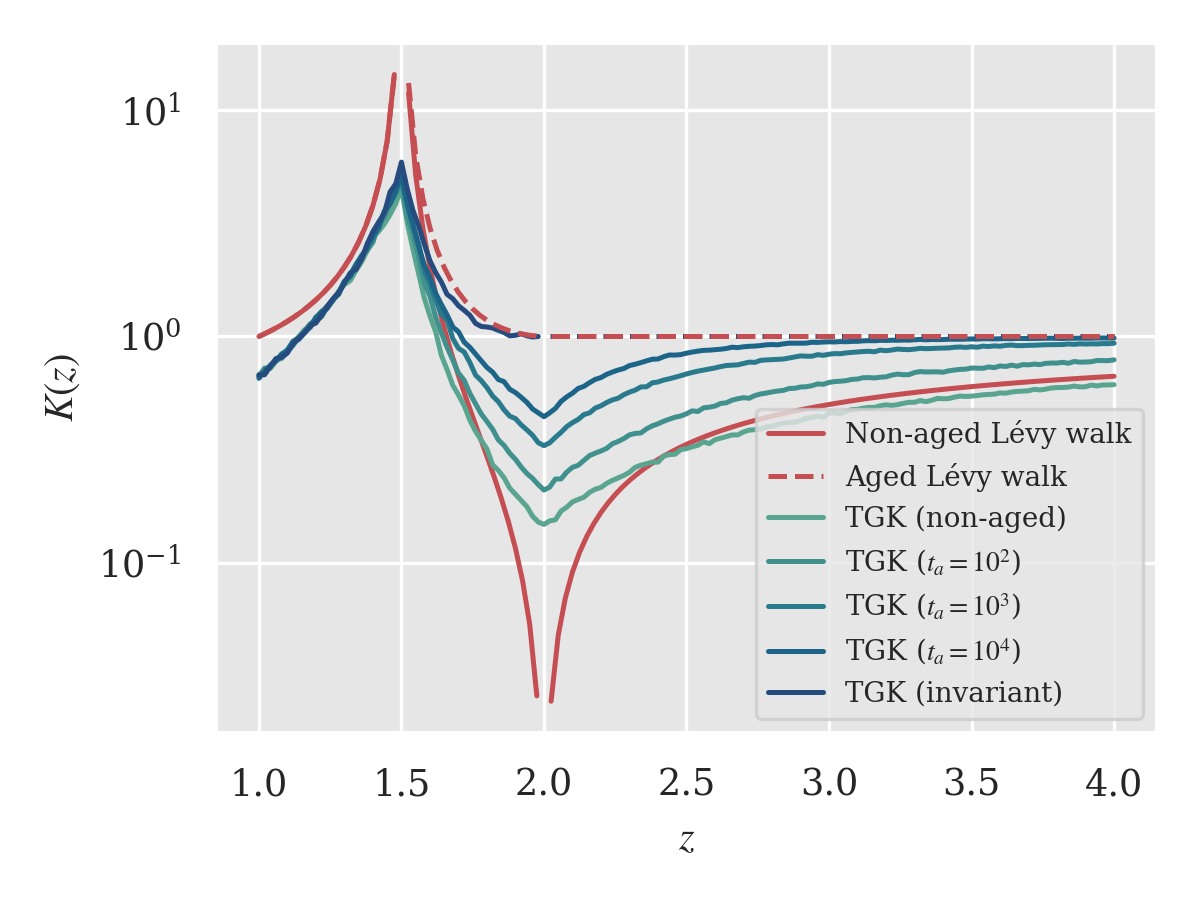}
\includegraphics[width=0.34\linewidth]{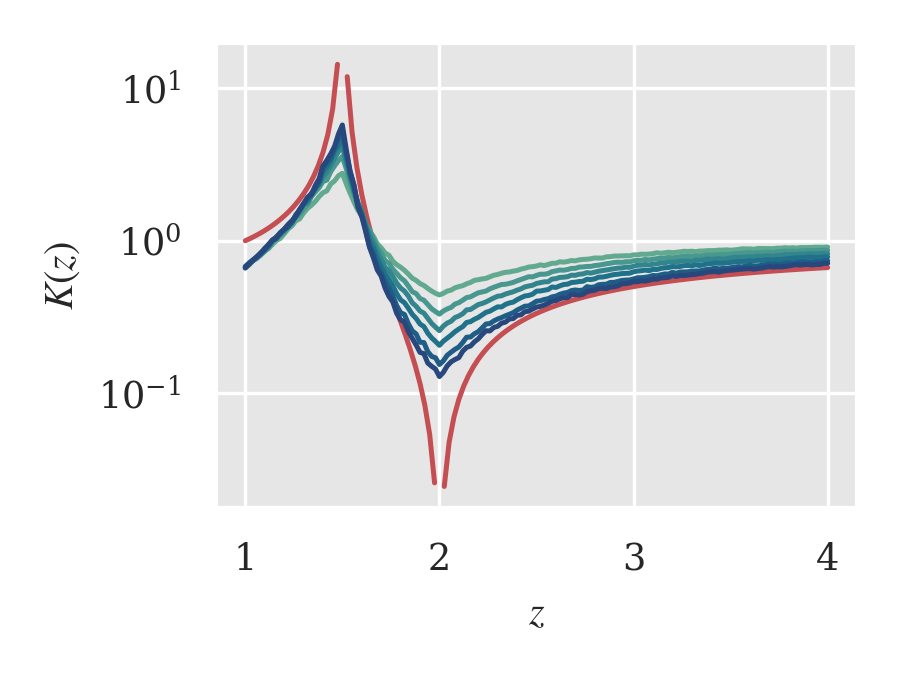}
\includegraphics[width=0.34\linewidth]{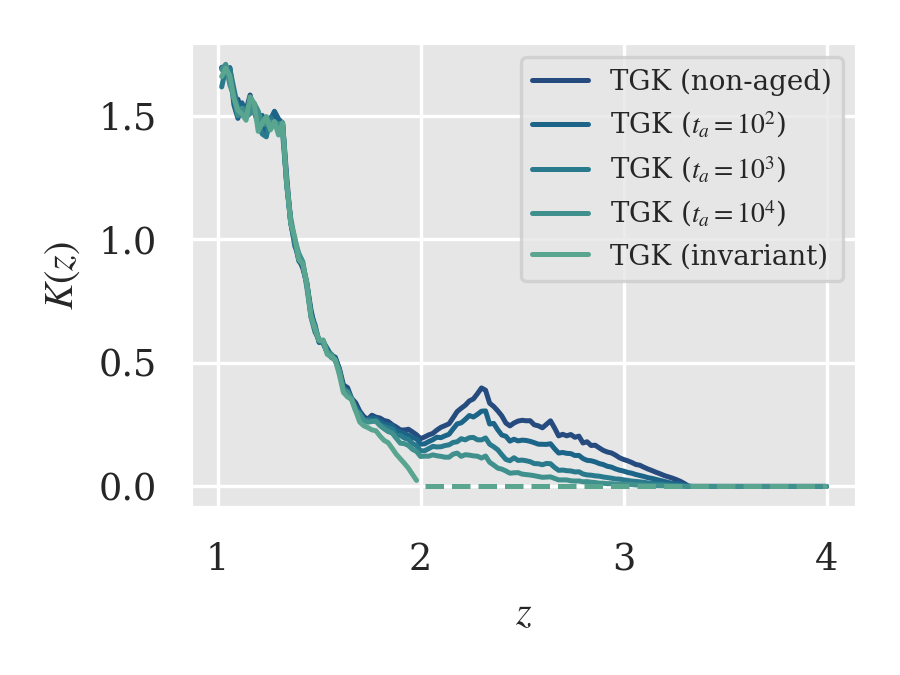}

\caption{Top: TGK approximations of $K(z)$ for the superdiffusive
  PM map, { fixing $R=1$,} starting from increasingly aged ensembles of particles
  initialised from a uniform density (aging duration increasing
  from light to dark), as well as starting from a normalisable
  invariant density (where it exists; darkest uppermost curve,
  blue online). Also shown are the results for the non-aged
  LW discussed above (smooth curve, red online) and
  the aged LW discussed in
  \cite{albers_exact_2018,albers_nonergodicity_2022}
  (dashed line, top, red online). All TGK plots are taken
  using $200$ terms of a TGK expansion from an ensemble of $10^4$
  simulated trajectories and consist of $151$ points ($50$ points
  in the case of the invariant ensemble). Bottom left: an identical
  plot to that shown in Fig.~\ref{fig:tgkconvfresh}, but with points
  initialised from an ensemble aged by $100$ timesteps. Bottom right:
  an identical plot to that shown above, for the subdiffusive extension
  of the map studied in \cite{korabel_fractal_2007}, mirroring Fig.~7
  in that paper, with $a=5$ fixed (aging duration increasing from dark
  to light, invariant curve shown at the bottom, in teal online).}
\label{fig:tgkaging}
\end{figure}

Finally, we explore the consequences of aging on numerical estimates
of $K(z)$. In the present setting, aging denotes the dependence of the
MSD, and hence of $K(z)$, on the duration, or aging time, between the
initialisation of the system dynamics and the start of its measurement
\cite{Bou92,Bar03,barkai_aging_2004,akimoto_aging_2013,metzler_anomalous_2014}.
Up to this point all estimates have been taken commencing from a
uniform density of $\{x_0\}$. However, many systems observed
in nature typically evolve for a long time before human measurement
of them can commence, and therefore we also consider the case where we
allow a degree of relaxation towards the invariant
density to take place before beginning measurements.
{We denote the number of timesteps during which the system is
allowed to evolve before measurement begins as the `aging time' $t_a$.
In such cases it is known that the dynamics of systems exhibiting aging
are characteristically dependent on the ratio of measurement time to
aging time, $t/t_a$ \cite{akimoto_aging_2013}.}

In Fig.~\ref{fig:tgkaging} we apply varying degrees of aging to an
ensemble of trajectories before calculating $K(z)$ via a {numerical}
TGK expansion.
{ We also show the same calculations starting from the invariant density of the reduced superdiffusive map where it exists for $z<2$, which is given in \cite{thaler_transformations_1983, akimoto_distributional_2012} to be}
\begin{equation} \label{eqn:invdensity}
\tilde{\rho}(x) \propto x^{1-z} + (1-x)^{1-z}
\end{equation}
{ (see also discussion in \cite{Zwei98, thaler_asymptotics_2000, korabel_separation_2010, akimoto_aging_2013}, where the invariant density of the reduced map \eqref{eqn:reduced} is shown to be asymptotic to $\tilde{\rho}(x) \propto x^{1-z}$).}
For $z\geq 2$ the only normalisable invariant density is the delta
function on $\{0,1\}$, which implies almost all particles undergo
purely ballistic motion in a single direction.  Also shown is the
corresponding analytic result for a LW in the limit of
large aging relative to measurement time, $t/t_a \ll 1$, as a special
case of recently reported results in \cite{albers_nonergodicity_2022}.
It is known that the MSD of aged walks corresponds exactly to the
ensemble-averaged time-averaged MSD, with the averaging time
equated to the aging time $t_a$ \cite{bothe_mean_2019}.
In \cite{albers_nonergodicity_2022} this was calculated for a
generalised LW with non-constant velocity, of
which our stochastic model is merely a special case.

The non-aged GDC from simulations is identical to that shown
in Fig.~\ref{fig:tgkconvfresh}. As the aging time increases, the
GDC from the aged ensemble converges uniformly both to the
GDC from the ensemble initialised on the invariant density,
where it exists (as the aged ensemble converges to the invariant one),
and to the aged LW results in \cite{albers_nonergodicity_2022},
shown in red. While for $z<\frac32$ in regime (I) the GDC is not visibly
affected by the aging, for $z>\frac32$ in regimes (II) and (III) the
qualitative effect of aging is easily observed -- most notably, where
in the non-aged scenario the dynamic transition at $z=2$ was marked by
a total vanishing of the GDC, aging seems to wipe out this phenomenon
completely, and the limit of small $t/t_a$ does not exhibit any at all
of the logarithmic corrections discussed previously at $z=2$ which
cause $K(z)$ to vanish.

The cause of this effect -- the uniform raising of the $K(z)$ curve
visible for $z>\frac32$ -- may be due to the fact that the aged densities
feature a more concentrated accumulation near the marginal points
$\{0,1\}$, which would lead to increased correlations in the short
term as more particles are in the ballistic phase of motion.
The disappearance of logarithmic corrections at $z=2$ may be explained by
the limiting density converging to the delta function at the point of
transition,
resulting in $K(z)\approx 1$.
Only once the measurement time greatly exceeds and dominates over the
aging time will the
the regular results be recovered
\cite{albers_nonergodicity_2022, akimoto_aging_2013}.

That there is no effect of aging on $K(z)$ in regime (I) for $z<\frac32$
is to be expected, since in this regime,
any initial density converges to the invariant one in exponential time. In
contrast, it is known that for $z\geq 2$ the PM map is only weakly ergodic
\cite{akimoto_aging_2013,metzler_anomalous_2014},
and therefore undergoes dynamical aging
\cite{akimoto_aging_2013,korabel_fractal_2007,
  Zwei98,thaler_transformations_1983}.
Interestingly though, some aging is observed on $K(z)$ even in the
intermediate superdiffusive regime (II), $\frac32 <z<2$, in which
{ the map is still ergodic, and
a normalisable invariant density still exists.} This reflects what was
described for LWs as `ultraweak' ergodicity breaking in that
parameter regime, which is characterised by a deviation in the GDC,
but not the exponent, of the time dependence of the MSD
\cite{metzler_anomalous_2014,froemberg_random_2013,froemberg_time-averaged_2013}.
{ The effect of varying initial conditions in regime II was also discussed in \cite{akimoto_distributional_2012} in relation to a similar superdiffusive Pomeau-Manneville-type map with bias, where a strong distributional response was identified on the MSD.}

A similar effect is observed for subdiffusive dynamics. In the bottom
right inset of Fig.~\ref{fig:tgkaging} is shown an equivalent plot showing
the GDC for the subdiffusive PM map used in \cite{korabel_fractal_2007},
with $a=5$, mirroring Fig.~7 in \cite{korabel_fractal_2007}.
The same phenomenon as in the superdiffusive case is observed:
above $z>\frac32$, aging begins to uniformly reduce $K(z)$ until it
converges to the result from the invariant density, which in this
case is $K(z)=0$ for $z>2$.

Incidentally, it was also observed in
\cite{KCKSG06,korabel_fractal_2007}
that for the subdiffusive map, there is strong numerical
evidence that the exact form of the GDC has a dependence
on the map's parameter which is fractal in nature, in contrast
to the smooth function generated by the LW; see
Fig.~\ref{fig:tgkaging} (bottom right) for an example.
It is already long known that fractal diffusion coefficients
are in fact quite typical for many classes of deterministic
dynamical systems
\cite{klages_microscopic_2007,KCKSG06, korabel_fractal_2007,
    korabel_fractality_2004, koza_fractal_2004}.
Although we do not conclude here whether or not this is true of the
superdiffusive extension given here, under variation of $z$,
the possibility cannot be excluded, which further precludes
an intimate comparison equating the map's dynamics to
those of the stochastic model.

We conclude this section by presenting a general principle:
that, for measurements taken over finite time, we believe it is
({perhaps} counterintuitively) preferable to average one's
ensembles over a uniform {initial} density
than from one closer to the more `natural' invariant
density. This is especially evident in { regime (III),
where $z>2$, for which the map is no longer ergodic, but also in
regime (II) where $z>\frac32$, where a strong dependence on initial
conditions is still observed.} We see that, this way, results
from a uniform initial density in practice
better reproduce the curve seen in the long measurement time limit,
without simulations becoming dominated by effects introduced by aging;
and for short sampling times, measuring from an aged or even invariant
density can produce {perverse} results which are not representative
of the long-measurement-time GDC of the process. Our exposition here
supplies some numerical evidence to {back} this principle.

\section{Conclusion\label{sec:concl}}

In this article we conducted an analysis on a superdiffusive extension
of the PM map. We calculated numerically the {MSD {exponent}
$\beta(R,z)$ and} GDC {$K(R,z)$} of the map, as
a function of the parameter of the map's nonlinearity {$z$, and its
coupling length $R$}, using {both direct simulations and} TGK
expansions, and compared these findings to a well-known CTRW model in
the form of a LW which models the map's dynamics
stochastically {in an important special case}.
{Within this special case $R=1$,} the LW model reproduces
the numerically calculated $K(z)$ well for some values {of the
parameter $z$,} but suffers a systematic and previously unexplained
and unaddressed quantitative deviation in the limit of normal
diffusion $z\to 1$. In this region a deeper investigation reveals the
dynamics are highly non-trivial,
in a much more subtle way than in the case of {the corresponding
subdiffusive extension}.
{Outside of this special case, the diffusive dynamics of the map
do not follow a LW, and indeed are shown not to be
superdiffusive at all, but instead are normally diffusive except in a
few key regions of the dual parameter space $(K,z)$ in which diffusion
is totally suppressed, due to dynamics becoming localised. Under
variation of these parameters in these regions, the dynamics undergo
a complex bifurcation scenario generating a fractal structure of stable
regions in phase space, which is produced numerically for the first time
and the contours of which are, to some extent, analytically calculated.}

We also investigated the effect of aging on our numerical findings.
We find, in accordance with recently reported results, that in the limit
of increased aging {time $t_a/t$}, the suppression {of} $K(z)$
at the transition from superdiffusion to ballistic motion is
eliminated, for finite measurement times, while the divergence at the
other, from normal to superdiffusion, is preserved.
This elimination is also reproduced in the case of the subdiffusive PM map.
We identify this as a problem of conflicting limits as both aging time and
measurement time go to infinity, which requires one to take caution
when conducting numerical or even physical experiments. We state a
general principle that uniform {initial} densities should be
preferred {in simulations} to more `natural' invariant ones if
{ adequately timely mixing} of the dynamics is not ensured, a claim which is
supported numerically.

In summary, our results illustrate again the subtlety to understand
diffusion generated by deterministic dynamical systems on the basis of
simple stochastic models that are applied {\em ad hoc} to a given
dynamical system. Often it seems taken for granted that the dynamics
of a `sufficiently chaotic' deterministic model boils down to a more
simplistic random walk-type process. Our analysis demonstrates, first
of all, that the celebrated matching of deterministic superdiffusion
in the lifted PM map to stochastic LWs only applies to a very
specific, singular subset of parameter values in the full parameter
space. In other words, the diffusive dynamics of this model turns out
to be highly unstable under parameter variation, affected by an
underlying basic bifurcation scenario. Secondly, for specific
parameter values where the map exhibits superdiffusion, as predicted
by stochastic LW theory, the parameter dependence of the GDC can
indeed to a large extent, {\em but by no means completely}, be
understood in terms of LWs generated by CTRW theory. This shows that
in general one may have to dig deeper in order to achieve a complete
understanding of the original deterministic dynamics. These results
may cast some doubt on the ubiquity of LWs in deterministic dynamical
systems and may be relevant when investigating more realistic systems
modeled by, for example, nonlinear deterministic equations of motion.

Correspondingly, it would be very interesting to test whether
dynamical transitions in the GDC, and associated aging effects, can be
observed in experiments. As candidates we would be thinking of
blinking quantum dots and cold atoms in atomic lattices, where
features of anomalous diffusion, aging, and weak ergodicity breaking
have already been reported \cite{SHB09}. It would also be worthwhile
to explore theoretically the parameter dependence of generalised
transport coefficients in other types of anomalous dynamical systems,
such as stochastic generalised Langevin dynamics \cite{CKW04} and
weakly chaotic models like polygonal billiards and related systems
\cite{klages_microscopic_2007}, which are known to generate a variety
of different dynamical transitions.  {From a mathematical angle, it is
  also an interesting open question as to whether the fractal nature
  of the map's bifurcation structure under variation of $R$ can be
  proven formally and not only shown numerically, for this and other
  related classes of diffusive maps.}

\appendix

\section{CTRW calculations}\label{appa}

In the superdiffusive extension of the map, the marginal point is mapped
one unit to the left or to the right (see Fig.~\ref{fig:extendmap}),
and therefore
in the CTRW model the time spent in the neighbourhood of the marginal
point is equated exactly with time spent in a ballistic motion with
constant velocity $v_0$
(we take $v_0 \equiv 1$) \cite{ShlKl85,zumofen_scale-invariant_1993},
\begin{equation} \label{eq:pt}
p(\ell \,|\, t) = \frac{1}{2}\,\delta(\abs{\ell} - v_0 t),
\end{equation}
where we enable the particle to move either left or right along a
one-dimensional axis with equal probability. These choices define the
model as a \emph{LW}
\cite{SKW82,KBS87,Shles87,zaburdaev_levy_2015}.
From \eqref{eq:wt} and \eqref{eq:pt} the joint density
\[ \psi(x,t) = w(t)\,p(x\,|\,t) \]
follows.


Under these assumptions, one may obtain the appropriate Montroll-Weiss
equation \cite{klafter_first_2011,
zumofen_scale-invariant_1993,zaburdaev_levy_2015},
\[ \hat{\tilde{P}}(k,s) = \Re\left\{ \frac{1-\tilde{w}(s+\im v_0 k)}{s+\im v_0 k} \right\}
  \frac{1}{1-\hat{\tilde{\psi}}(k,s)}, \]
where $P(x,t)$ is the spatial density of the random walker as a
function of time, and $(k,s)$ are the respective Fourier- and
Laplace-transformed variables for $(x,t)$
(such transformations are henceforth indicated
by \ $\hat{}$ \ and \ $\tilde{}$ \ respectively).
The Laplace-transformed MSD can then be found very easily by
\begin{equation} \label{eq:x2}
\langle \tilde{x}^2(s) \rangle = \left.\pdv[2]{}{k}\Pks\right|_{k=0}
\end{equation}
to which an inverse transform may be performed to obtain
$\langle x^2 \rangle (t)$.
Frequently such inverse transforms are intractable and instead
we rely on asymptotic analysis as $(k,s)\to (0,0)$ to obtain the
desired quantities to the required order.
The GDC \cite{metzler_random_2000,KCKSG06,korabel_fractal_2007}
can then be defined by
\begin{equation} \label{eq:Kz}
K(z) := \lim_{t\to\infty} \frac{\langle x^2 \rangle}{t^\beta}
\end{equation}
for the appropriate $\beta>0$ wherever this limit exists, and is finite
and non-zero.

\section{Bifurcation calculations}\label{appb}

{We restrict our discussion to the interval $[-\frac12, \frac12]$.
Let $M^+(x)$ and $M^-(x)$ denote the variants of the map to be
applied when $x$ is positive or negative respectively,
ie.,\ from \eqref{eq:pm},
\begin{align*}
M^+(x) &= x + R((2x)^z - 1), \\
M^-(x) &= -M^+(-x) \\ &= x - R((-2x)^z - 1),
\end{align*}
which for $z=1$ reduces to
\begin{align*}
M^+(x) &= x + 2Rx - R, \\
M^-(x) &= x + 2Rx + R,
\end{align*}
and for $z=2$
\begin{align*}
M^+(x) &= x + 4Rx^2 - R, \\
M^-(x) &= x - 4Rx^2 + R.
\end{align*}}
{For $z=1$, it is apparent by inspection that the box onto which
the dynamics are localised is given by
\[ [M^+(0), M^-(0)] = [-R, R]. \]
Therefore let us restrict ourselves further to the interval $[-R, R]$.
Inspecting $M^2$, we identify an invariant region at
$[-x_\alpha, x_\alpha]$ with
\begin{align*}
x_\alpha &= M^+(M^-(0)) \\
&= \left. (x+2Rx+R)+2R(x+2Rx+R)-R \right|_{x=0} \\
&= \left. x(1+4R+4R^2)+2R^2 \right|_{x=0} \\
&= 2R^2
\end{align*}
and at $[-R, -x_\beta] \cup [x_\beta, R]$ with
\begin{align*}
x_\beta &= M^2(-R) = M^-(M^-(-R)) \\
&= \left. (x+2Rx+R)+2R(x+2Rx+R)+R \right|_{x=-R} \\
&= \left. x(1+4R+4R^2)+2R^2+2R \right|_{x=-R} \\
&= R-2R^2-4R^3.
\end{align*}
The unstable fixed point $x_u>0$ (and identically $-x_u<0$) appears when
$M^2(x_u) = M^-(M^+(x_u)) = x_u$, ie.\
\begin{gather*}
(x_u+2Rx_u-R)+2R(x_u+2Rx_u-R)+R = x_u \\
\implies x_u(1+4R+4R^2)-2R^2 = x_u \\
\implies x_u(4R+4R^2) = 2R^2 \\
\implies x_u = \frac{R^2}{2R(1+R)}.
\end{gather*}
It can be shown that all three points coincide at $R_c$ given by the sole
positive real solution to the cubic equation
\begin{align*}
R_c-4R_c^2-4R_c^3 &= 0 \\
\implies R_c &= -\frac12 + \sqrt{\frac12} \approx 0.207, \\
x_c = x(R_c) &= \frac32 - \sqrt{2} \approx 0.0858.
\end{align*}}
{For $z=2$, let us pre-calculate
\begin{align*}
(x-4Rx^2+R)^2 &= x^2+16R^2x^4+R^2 \\
&\qquad \phantom{\pi}-8Rx^3+2Rx-8R^2x^2
\end{align*}
which gives
\begin{align*}
&\left. (x-4Rx^2+R)^2 \right|_{x=0} = R^2; \\
&\left. (x-4Rx^2+R)^2 \right|_{x=-R} = 16 R^6.
\end{align*}
Then we calculate
\begin{align*}
x_\alpha &= M^+(M^-(0)) \\
&= \left. (x-4Rx^2+R)+4R(x-4Rx^2+R)^2-R \right|_{x=0} \\
&= \left. (x-4Rx^2+R)+4R[R^2]-R \right|_{x=0} \\
&= 4R^3
\end{align*}
and
\begin{align*}
x_\beta &= M^2(-R) = M^-(M^-(-R)) \\
&= \left. (x-4Rx^2+R)-4R(x-4Rx^2+R)^2+R \right|_{x=-R} \\
&= \left. (x-4Rx^2+R)-4R[16R^6]+R \right|_{x=-R} \\
&= R -4R^3 -64R^7.
\end{align*}
Then $x_u>0$ appears at
$M^2(x_u) = M^-(M^+(x_u)) = x_u$, ie.\
\begin{gather*}
(x_u+4Rx_u^2-R)-4R(x_u+4Rx_u^2-R)^2+R = x_u \\
\implies (-64R^3)x_u^4 + (-32R^2)x_u^3 + (32R^3)x_u^2 \\
\qquad\qquad\qquad + (8R^2)x_u + (-4R^3-R) = 0 \\
\implies x_u = \frac{-1+\sqrt{1+4R^2}}{4R}.
\end{gather*}
Note that $x_u$ is the solution of a quartic equation whose coefficients
vary in $R$; hence for increasing $z$ these quantities become very difficult
to calculate by hand.
The critical point $R_c$ at which these coincide is given by the sole
real positive solution to a polynomial equation of degree 7, which we
solve here only numerically:
\begin{align*}
R_c-8R_c^3-64R_c^7 &= 0 \\
\implies R_c &\approx 0.337, \\
x_c &\approx 0.153.
\end{align*}
Extending these calculations to higher, or to non-integer, values of
$z$ would undoubtedly lead to quite unwieldy equations, hence values
of $z\in\{1,2\}$ were selected here, although the qualitative
dynamics can be seen numerically to be similar.}

\begin{figure}
\centering
\includegraphics[width=0.8\linewidth]{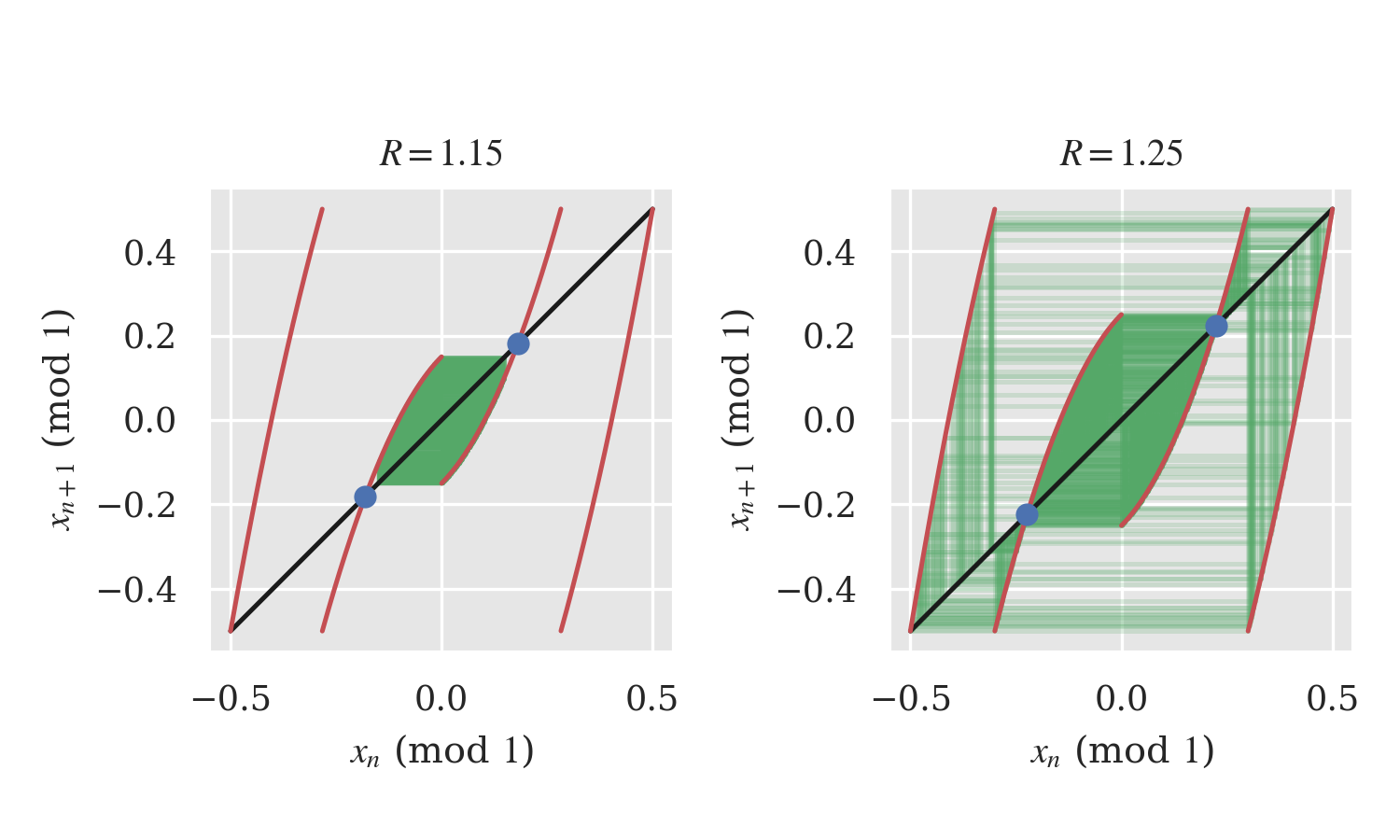}
\caption{Plots of sample trajectories in the Pomeau-Manneville map
(mod $1$), over the region $[-\frac12, \frac12]$,
for $z=2$, $R\in\{1.15, 1.25\}$. The repelling region of the
phase space becomes accessible as $R$ exceeds $R_d \approx 1.207$, see text.
The critical points $\pm x_d$ are shown in blue.}
\label{fig:rc}
\end{figure}

In Fig.~\ref{fig:rc}, we see that the repelling region of phase space becomes
accessible when the branch edge (located near $x_n=0$) exceeds the critical
point $x_d$, which is a (mod $1$) repelling fixed point. In the full map, this is the point where $M(x) = x\pm 1$. Precisely, when $z=2$, this is the point
\begin{align*}
M^+(x_d) = x_d + 4Rx_d^2 - R &= x_d - 1 \\
\implies x_d &= \sqrt{\frac{R-1}{4R}}.
\end{align*}
The critical value $R_d$ is the value of $R$ at which $M^-(0)-1$
first exceeds $x_d$, ie.,
\begin{gather*}
M^-(0)-1 = R_d-1 = \sqrt{\frac{R_d-1}{4R_d}} = x_d \\
\implies 4R_d^2 - 4R_d + 1 = 0 \\
\implies R_d = \frac{1+\sqrt{2}}{2} \approx 1.207.
\end{gather*}
Note that this is remarkably close to, but does not quite fit with,
the numerically identified values for $\mathcal{N}_1(z)$
(the values of $R$ at which normal diffusion is restored)
depicted in Fig.~\ref{fig:schulz_beta}. Whether this is due to
artefacts in the numerical simulations, or to something more subtle
dynamically, remains an open question.

\begin{figure*}
\centering
\includegraphics[width=0.95\linewidth]{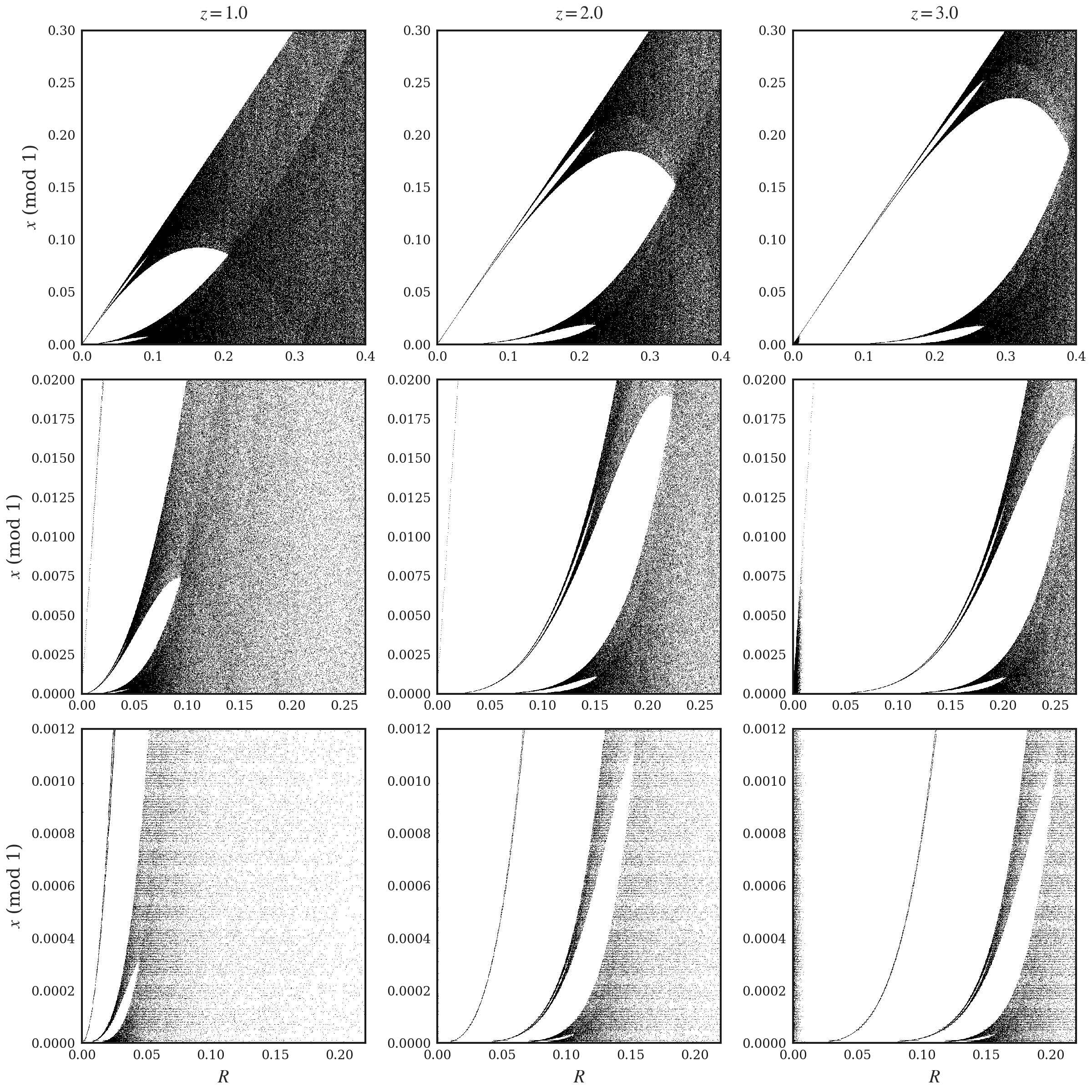}
\caption{Zooms of Fig.~\ref{fig:schulz_main} for $z\in\{1,2,3\}$
(left to right), for three varying levels of zooming (top to bottom),
indicating the hierarchy of bubble-like structures in the neighbourhood
of $R=0$, $x=0$. We conjecture this hierarchy to be infinite.}
\label{fig:schulz_large}
\end{figure*}


%

\end{document}